%% file: Main.tex
\newcommand{\sysname}{QuickCue}

\DocumentMetadata{}
\documentclass[sigconf]{acmart} 
\usepackage{balance}  
\usepackage{adjustbox}

\usepackage{geometry} 
\usepackage{changepage} 
\usepackage{tcolorbox} 
\usepackage{mdframed} 
\usepackage{xcolor}
\usepackage{placeins}
\usepackage{graphicx}
\usepackage{array}
\usepackage{amsthm}
\usepackage{amsmath,amsfonts}
\usepackage{algorithmic}
\usepackage{booktabs}
\usepackage{hyperref}
\usepackage{textcomp}
\usepackage{multirow}
\usepackage{array}
\usepackage{xcolor}
\usepackage{enumitem}
\usepackage{makecell}
\usepackage{multicol}
\usepackage{graphicx}
\usepackage{layouts}
\usepackage{booktabs}
\usepackage{dcolumn}
\usepackage{colortbl}
\usepackage[T1]{fontenc}
\usepackage{subcaption}
\usepackage{makecell}
\usepackage{amsmath}
\usepackage{bm}

\definecolor{tableline-gray}{gray}{0.7}
\definecolor{light-gray}{gray}{0.9}
\definecolor{word-violet}{RGB}{113,137,191}

\AtBeginDocument{%
  }

\setcopyright{acmlicensed}
\copyrightyear{2025}
\acmYear{2025}
\acmDOI{XXXXXXX.XXXXXXX}
\acmConference[W4A '25]{Make sure to enter the correct
  conference title from your rights confirmation email}{April 28-29,
  2025}{Sydney, Australia}

\begin{document}

\title{Adapting Online Customer Reviews for Blind Users: A Case Study of Restaurant Reviews}




\author{Mohan Sunkara}
\orcid{0000-0002-6970-0203}
\affiliation{%
  \institution{Old Dominion University}
  \department{Department of Computer Science}
  \city{Norfolk}
  \state{Virginia}
  \country{USA}
}
\email{msunk001@odu.edu}

\author{Akshay Kolgar Nayak}
\affiliation{%
  \institution{Old Dominion University}
  \department{Department of Computer Science}
  \city{Norfolk}
  \state{Virginia}
  \country{USA}
}
\email{anaya001@odu.edu}

\author{Sandeep Kalari}
\affiliation{%
  \institution{Old Dominion University}
  \department{Department of Computer Science}
  \city{Norfolk}
  \state{Virginia}
  \country{USA}
}
\email{skala003@odu.edu}



\author{Yash Prakash}
\affiliation{%
  \institution{Old Dominion University}
  \department{Department of Computer Science}
  \city{Norfolk}
  \state{Virginia}
  \country{USA}
}
\email{yprak001@odu.edu}

\author{Sampath Jayarathna}

\affiliation{%
  \institution{Old Dominion University}
  \department{Department of Computer Science}
  \city{Norfolk}
  \state{Virginia}
  \country{USA}
}
\email{sampath@cs.odu.edu}

\author{Hae-Na Lee}

\affiliation{%
  \institution{Michigan State University}
  \department{Department of Computer Science and Engineering}
  \city{East Lansing}
  \state{Michigan}
  \country{USA}
}
\email{leehaena@msu.edu}

\author{Vikas Ashok}

\affiliation{%
  \institution{Old Dominion University}
  \department{Department of Computer Science}
  \city{Norfolk}
  \state{Virginia}
  \country{USA}
}
\email{vganjigu@odu.edu}

\renewcommand{\shortauthors}{Mohan Sunkara et al.}

\input{Sections/0.Abstract}




\maketitle

\input{Sections/1.Introduction}
\input{Sections/2.RelatedWork}

\input{Sections/3.InterviewStudy.tex}

\input{Sections/4.Architecture}

\input{Sections/5.Evaluation}
\input{Sections/6.Discussion}
\input{Sections/7.Conclusion.tex}
\section*{Acknowledgments}
We sincerely thank Md Javedul Ferdous, Nithiya Venkatraman, and Abhinav Sai Choudary Panchumarthi for their assistance in preparing the seminal version of the manuscript.


\balance

\bibliographystyle{ACM-Reference-Format}
\bibliography{Main}



\end{document}

%% file: Sections/0.Abstract.tex
\begin{abstract}
Online reviews have become an integral aspect of consumer decision-making on e-commerce websites, especially in the restaurant industry. Unlike sighted users who can visually skim through the reviews, perusing reviews remains challenging for blind users, who rely on screen reader assistive technology that supports predominantly one-dimensional narration of content via keyboard shortcuts. In an interview study, we uncovered numerous pain points of blind screen reader users with online restaurant reviews, notably, the listening fatigue and frustration after going through only the first few reviews. 
To address these issues, we developed \sysname{} assistive tool that performs aspect-focused sentiment-driven summarization to reorganize the information in the reviews into an alternative, thematically-organized presentation that is conveniently perusable with a screen reader. At its core, \sysname{} utilizes a large language model to perform aspect-based joint classification for grouping reviews, followed by focused summarizations within the groups to generate concise representations of reviewers' opinions, which are then presented to the screen reader users via an accessible interface.
Evaluation of \sysname{} in a user study with $10$ participants showed significant improvements in overall usability and task workload compared to the status quo screen reader.
\end{abstract}

%% file: Sections/1.Introduction.tex
\section{Introduction}
Online reviews have become a cornerstone of modern consumer decision-making, offering valuable insights into products, services, and experiences~\cite{Karamad2023THEIO, Racherla2012PerceivedO, Yayli2012eWOMTE}. This has especially been the case in the restaurant industry, with reviews and ratings providing diners with information about food quality, ambiance, and service, thereby helping them make informed choices~\cite{Luca2016ReviewsRA, Fang2022TheEO}. Therefore, the presentation of information in user reviews must be as holistic and fair as possible, to avoid inducing consumer biases and harming a restaurant's reputation. While present applications (e.g., Google Maps) do include an assortment of features in their interfaces to help prospective diners make fully-informed decisions, these features are presently insufficient for blind users who rely on screen reader assistive technology to interact with these applications. 

\begin{figure*}[t!]   
    \centering 
    \includegraphics[width=\textwidth]{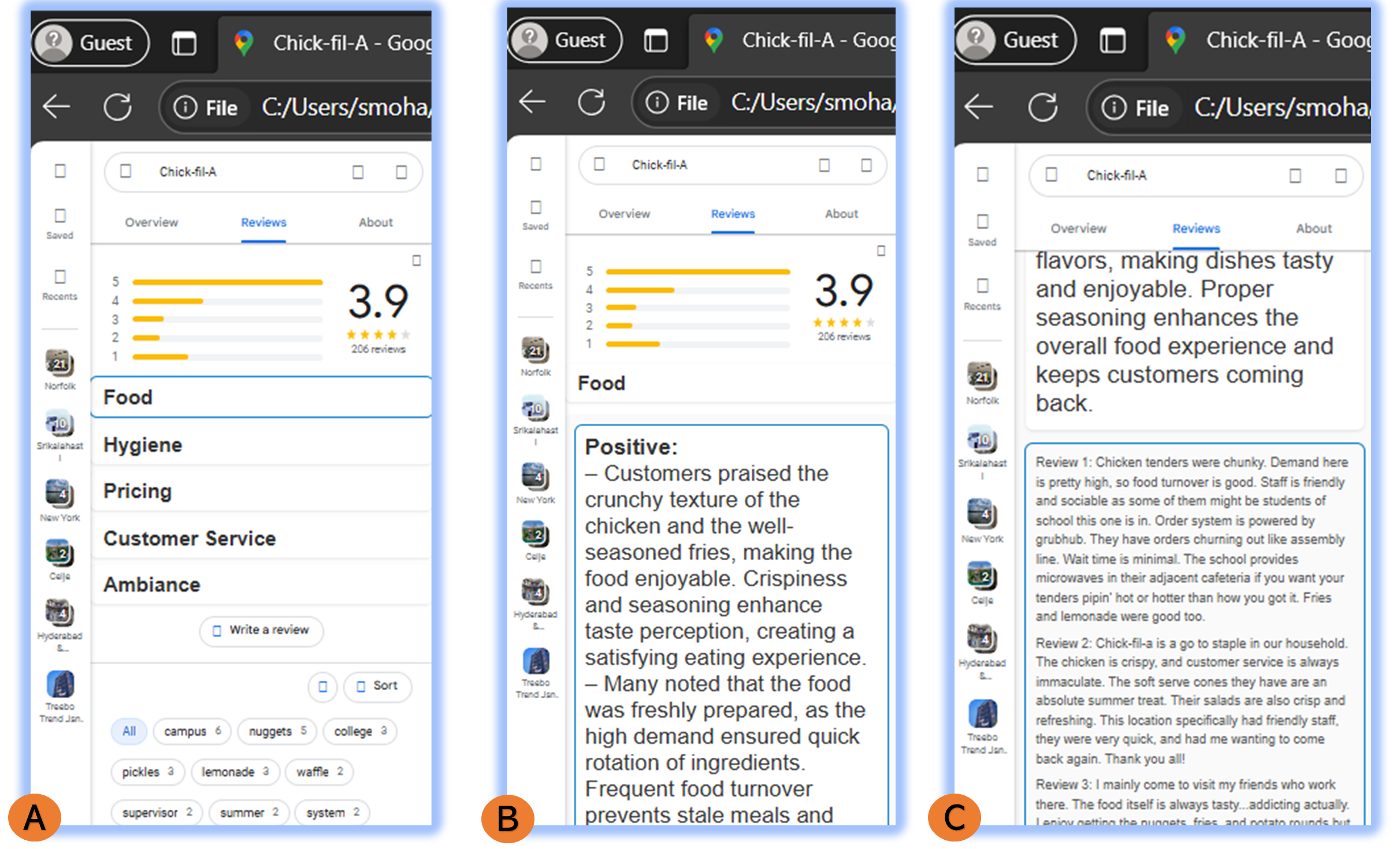}
    \caption{\sysname{}'s interface hierarchy: (A) displays the default list of five aspects, (B) shows the positives and negatives of each aspect, and (C) shows the original reviews associated with each aspect-sentiment pair.}
    \label{fig:intro}
    \Description{Figure shows three side-by-side screenshots from a desktop Google Maps interface as updated by QuickCue. The first screenshot displays a list of aspects (for example, Food, Hygiene, Customer Service, and Ambiance). The second screenshot presents positive and negative summaries for the selected aspect, while the third screenshot reveals a set of individual reviews that expand upon a chosen aspect-sentiment pair.}
\end{figure*}

A screen reader (e.g., JAWS~\cite{jaws}, NVDA~\cite{nvaccess}, VoiceOver~\cite{voiceover}) narrates web-application content based on the order in the webpage's document object model (DOM), essentially enforcing a one-dimensional interaction with the content. Although a screen reader offers numerous keyboard shortcuts to aid navigation, the ``press-and-listen'' paradigm inherently limits the efficiency and usability of accessing and consuming information, including user reviews, in websites~\cite{prakash2024all,lazar2007frustrates}. In contrast, sighted users can effortlessly scan and skim online reviews, by leveraging visual cues to quickly pinpoint relevant content. In an interview study with $30$ blind screen reader users, we found that participants often experienced listening fatigue after perusing only a few reviews, they faced difficulties in finding specific information, and in general they struggled with navigating unstructured and repetitive reviews. Almost all participants expressed a need for an alternative thematic presentation of reviews (i.e., grouping according to food quality, ambiance, service, etc.), with further bifurcation of information along the lines of `positives' and `negatives' (e.g., good/bad experiences about service, food, or ambiance). 

To address the aforementioned interaction challenges as well as adapt the presentation of information to match users' needs and preferences, we developed \sysname{}, a novel assistive tool embodied as a browser extension, that augments the existing interface (see Figure~\ref{fig:intro}) with additional content comprising aspect-based organization of information mined from reviews. To generate such as an alternative presentation, \sysname{} performs the following two core tasks: (i) Joint classification of reviews to determine the aspect-sentiment pairs covered in the reviews; followed by (ii) Focused summarization of select reviews associated with each aspect-sentiment pair. \sysname{} performs both these tasks by leveraging the GPT-4 large large model (LLM), specifically using the \textit{clue and reasoning prompt engineering} strategy~\cite{sun-etal-2023-text} for joint classification, and \textit{directed stimulus prompt engineering} strategy~\cite{li2024guiding} for focused summarization of select reviews. The user interface, as shown in Figure~\ref{fig:intro}, hierarchically presents the generated information, with only aspects listed in the first level~(Figure~\ref{fig:intro}(A)), to the positive and negative summaries in the second level~(Figure~\ref{fig:intro}(B)), to the subset of reviews associated with the aspect-sentiment pair in the last level~(Figure~\ref{fig:intro}(C)). 

A user study with $10$ blind participants showed that \sysname{} significantly improved usability and reduced interaction workload for blind screen reader users while perusing user reviews, compared to the status quo. Furthermore, a majority of the participants stated that \sysname{} would enable them to make more informed decisions regarding choice of restaurants.
In sum, this paper makes the following contributions:
\begin{itemize} 
\item Interview study with $30$ blind users uncovering their interaction challenges and needs regarding online reviews.
\item The design, development, and evaluation of \sysname{}, a novel assistive tool for blind users to efficiently consume customer reviews. 
\end{itemize}

%% file: Sections/2.RelatedWork.tex
\section{Related Work}

\subsection{Online Review Systems}

Extensive research has explored the impact of online reviews on consumer decision making~\cite{liu2022sentiment,siahaan2017new,dash2021personalized,sparks2011impact}. 
Liu et al.~\cite{liu2022sentiment} found that online reviews allow consumers to make more informed choices, acting as digital word-of-mouth. The impact of reviews extends to sales performance, as even slight changes in ratings can lead to substantial variations in revenue~\cite{luca2016reviews}. Credibility is a key factor in the effectiveness of online reviews, with detailed, balanced, and authentic feedback being perceived as more trustworthy~\cite{filieri2015makes}. 

In the restaurant domain, consumer reviews typically include detailed feedback on food quality, service, ambiance, and price, all of which are key factors influencing a prospective diner's restaurant selection. Campos et al.~\cite{siahaan2017new} applied natural language processing techniques to analyze sentiments in online reviews, showing how positive or negative sentiments directly impact restaurant reputation and customer expectations. Dash et al.~\cite{dash2021personalized} further extended this work by demonstrating the effectiveness of deep learning models in extracting relevant attributes from online reviews to recommend dishes, highlight popular items, and even tailor experiences based on individual customer preferences. The above research works underline the critical role of online reviews in shaping consumer perceptions and behaviors. However, these works do not explore users' interaction and engagement with the content in reviews. While online review systems play a crucial role in shaping consumer preferences, blind screen reader users are presently unable to fully exploit these systems, as the present user interfaces are mostly designed for visual interaction. We address this issue in this paper by building \sysname{} that dynamically augments the current review systems with a screen-reader friendly interface to conveniently peruse information in reviews.

\subsection{Web Interaction with Screen Readers}
As mentioned earlier, blind people interact with digital applications, including web applications, using screen-reader assistive technology such as JAWS, NVDA, or VoiceOver. A screen reader transforms the two-dimensional graphical interface of a web page into a linear, one-dimensional list of on-screen elements (such as headers, text, buttons, and menus) for auditory navigation. This sequential press-and-listen method of navigation has been found to create significant accessibility and usability challenges for blind users~\cite{abuaddous2016web,oh2021image,billah2017speed,ashok2014wizard,ashok2017web,melnyk2015look,guerino2020usability,zong2022rich}, despite the availability of several accessibility guidelines~\cite{caldwell2008web,world1999web,chisholm2001web} and accessibility checking-aids for web developers~\cite{abascal2019tools,brajnik2004comparing,kumar2021comparing,abou2008web}.

While the accessibility challenges have been extensively investigated in prior works~\cite{kherwa2017latent,low2019twitter,wu2017automatic}, relatively fewer studies have focused on the usability of web interaction for screen reader users~\cite{aydin2020sail,ferdous2022insupport,lee2022customizable}. Usability, i.e., the ease, efficiency and satisfaction with which users can accomplish tasks, is equally important in web interaction for blind users, with many studies showing that screen reader users are typically an order of magnitude slower than sighted peers in doing the same web tasks~\cite{do2019comparing,luy2021toolkit}. While usability-enhancing solutions for blind users have been proposed in the literature~\cite{prakash2024all, sunkara2023enabling, lee2022customizable, prakash2023autodesc,ferdous2022insupport, ferdous2021semantic,lee2021towards, lee2020rotate,ashok2019auto,khanna2024hand, fakrudeen2017finger, brock2015interactivity}, these have predominantly focused on the general efficiency of webpage navigation, and as such they are inadequate in their ability to address domain-specific challenges involved in online review systems. Review systems are not only text-heavy with significant information redundancy, but often require nuanced understanding of tone, sentiment, and other specific preferences or features that traditional screen readers and other extant usability solutions are currently unable to support for assisting blind users. A tailored solution is therefore needed to address this issue and make online review systems more usable for blind screen reader users, which is the focus of this paper.

\subsection{User Interfaces of Review Systems}
The effective organization of information plays a crucial role in enhancing user experience, especially in text-heavy online review systems. Therefore, prior works have looked into methods such as clustering, sentiment-based categorization, and personalized filtering to structure data in reviews into more digestible formats~\cite{khalid2020gbsvm,khan2020sentiment,xu2012personalized}. For instance, faceted navigation has been widely applied in e-commerce platforms to allow users to filter reviews based on specific attributes, such as taste, portion size, and service quality~\cite{vandic2024framework}. Similarly, researchers have explored sentiment-based organization of restaurant reviews, finding that customers tend to prioritize attributes such as food quality and pricing when assessing menu items~\cite{zuheros2021sentiment,ara2020understanding}. However, these works have all focused mostly on sighted-user interaction, and as such do not fully address the unique needs of blind screen reader users.

Another factor to consider while presenting reviews to blind users is the redundancy of information. As screen reader interaction consumes significant time and effort ~\cite{lazar2007frustrates,sharif2021understanding}, repetitive reviews with little new information can be burdensome for blind users, since they are unable to quickly skim through the reviews like their sighted counterparts. Text summarization ~\cite{el2021automatic,tas2007survey}, therefore, can be a valuable tool to address this issue. Especially, the recent large language models such as GPT-4~\cite{achiam2023gpt}, Gemini~\cite{team2023gemini}, and Llama~\cite{touvron2023llama} which have demonstrated remarkable summarization capabilities across diverse domains with either prompting ~\cite{srinivas2024evaluation,chhikara2024lamsum,jin2023binary} or minimal fine-tuning~\cite{yang2023exploring}, can be very useful to compact information in reviews across different aspects and granularity before providing them to the blind users. However, ensuring factual accuracy and maintaining relevance in generated summaries requires domain-specific adaptation (e.g., through tailored few-shot prompts) and is currently an active area of research~\cite{frincu2023search,wallace2021generating}. While some applications, e.g., Google Maps, are leveraging these models to summarize information in reviews, these are mostly `high-level' short summaries capturing multiple aspects; these applications do not provide specific summaries pertaining to individual aspects, which can be more informative to users.

%% file: Sections/3.InterviewStudy.tex
\section{Interview Study}
We conducted an IRB-approved interview study with 30 screen reader users to investigate their needs and challenges regarding online customer reviews, particularly in the restaurant domain. The participants were recruited via email lists and word-of-mouth. The average age of the participants was $43.2$ years (median: 44, min: 22, max: 63), and the gender distribution was 13 male and 17 female. The inclusion criteria were: (i) Proficiency in web screen reading; (ii) Familiarity with online review systems including restaurant reviews; and (iii) Proficiency in English. The interviews were semi-structured to allow deeper discussions into the participants' needs and issues, with each discussion initiated by a set of careful crafted seed questions. Examples of seed questions included: \textit{What are the primary challenges you encounter while navigating restaurant reviews?}, \textit{What key information do you prioritize when browsing reviews?},  \textit{How do you typically search for the key details in the reviews?}, and \textit{What improvements would make it easier for you to find relevant and helpful information in restaurant reviews?}. The interview feedback was qualitatively analyzed using the standard open coding and axial coding methods~\cite{saldana2021coding}, where we iteratively examined the transcribed interview data to identify key insights and pain points. 

\subsection{Findings}

\noindent \textbf{Information overload and listening fatigue.}
A majority ($26$) of the participants reported struggling with navigating large volumes of reviews, describing it as a significant challenge due to repetitive and redundant content that caused frustration and fatigue. In this regard, one participant P8 stated, ``It’s frustrating to go through 10 reviews that say the same thing -- great food, nice ambiance. I need more details, like whether the restaurant can accommodate dietary restrictions or if the seating is comfortable and accessible.'' Another participant P4 shared a similar sentiment, ``Sometimes I just give up because the information feels repetitive and boring.'' More than half ($17$) of the participants expressed that often stopped perusing reviews after listening to only the first few reviews due to listening fatigue. Eleven participants further felt that the present interface was not `fair' to them in this regard, as they could not obtain sufficient information to make informed decisions. Towards this, one participant P12 mentioned, ``I cannot go beyond 4 to 5 comments without getting tired ... sometimes even less if the first couple of reviews are too long ... so I don't get the full picture of what is good and what is bad ... just have to decide based on opinions of a couple of folks, which is obviously unfair.'' 

\noindent \textbf{Outdated information and reviews.} All participants mentioned that they often came across reviews that contained outdated information about the different aspects of the restaurants. For examples, P6 said, ``I first check the menu and then look at the reviews to see which items are good. But some of the reviews don't make any sense, since they mention dishes that don't exist in the menu ... perhaps the menu has been changed after the review was written, who knows''. Another participant P23 echoed, ``A lot of things changed after COVID ... many reviews before COVID are no longer useful.'' When probed regarding their preferred time threshold for reviews, 17 participants mentioned that they wished to only view reviews in the past 1 year, 7 participants indicated two years as their preferred limit, whereas the remaining 6 participants mentioned that they were only interested in reviews from the past six months. In this regard, one participant P17 stated, ``Restaurant staff, food quality, and service keep changing all the time. What was good a few years ago may not be good anymore ... some of the bad stuff might have also improved over time.''

\noindent \textbf{Thematic organization of information in reviews.}
Most ($28$) of the participants expressed a need for an alternative aspect-based (e.g., food quality, hygiene, and ambiance) organization of information in reviews, as they felt that this type of organization helpful because it allowed them to quickly identify the aspects most relevant to their needs. Sixteen participants further suggested summarizing reviews within each aspect to avoid redundancy in the content and also drown out vague uninformative reviews (e.g., \textit{good food!}). As for the preferred aspects, food quality and pricing were unsurprisingly specified as high priority (by $26$ and $22$ participants respectively). The customer service, hygiene, and ambiance aspects were also mentioned as important by a sizable chunk of the participant pool ($19$, $16$, and $12$ participants respectively). For instance, P5 stated, ``I attend for the food; however, if the environment is noisy or cramped, it detracts from the overall experience.'' Another participant P27 reiterated this perspective: ``Good service is essential. A rude waiter can make even a tasty meal forgettable.''

\noindent \textbf{Sentiment-based insights.}
Nearly two-thirds ($19$) of the participants expressed a desire for sentiment-based segregation of information in the reviews. For instance, P16 said, `` I simply like to know what is good and what is bad. What is nice about the food... which dishes to avoid.. is the place too crowded on the weekends... are the prices reasonable. If I can easily get this information without having to search for it myself, it will save me a lot of time.''  The preferences for the `positives' vs. `negatives' however varied across the different aspects. While some of the participants were more interested in the positive feedback regarding food quality (e.g., `What items are most recommended here?' -- P14), others were more interested in the negative feedback regarding hygiene (e.g., `Do they have a hand sanitizer at the entrance?' -- P23). Similarly, regarding the pricing, customer service, and ambiance, the participants leaned more towards negative, negative, and positive feedback respectively. 

\noindent \textbf{Summary.} The participants pointed out several interaction issues with the current presentation of customer reviews, including listening fatigue, frustration after listening to only a few reviews, content redundancy, and difficulty in searching for specific information. To address these issues, most participants suggested an alternative presentation of information in the customer reviews, specifically along the concepts of themes (or aspects) and customer sentiment. With Google Maps as the vehicle for our investigation, we developed \sysname{} prototype that generates such an alternative screen reader-friendly thematic presentation of information in restaurant reviews, as described next.

%% file: Sections/4.Architecture.tex
\section{\sysname{} Prototype Design}

\subsection{Overview}

\begin{figure*}[t!]   
    \centering 
    \includegraphics[width=0.99\textwidth]{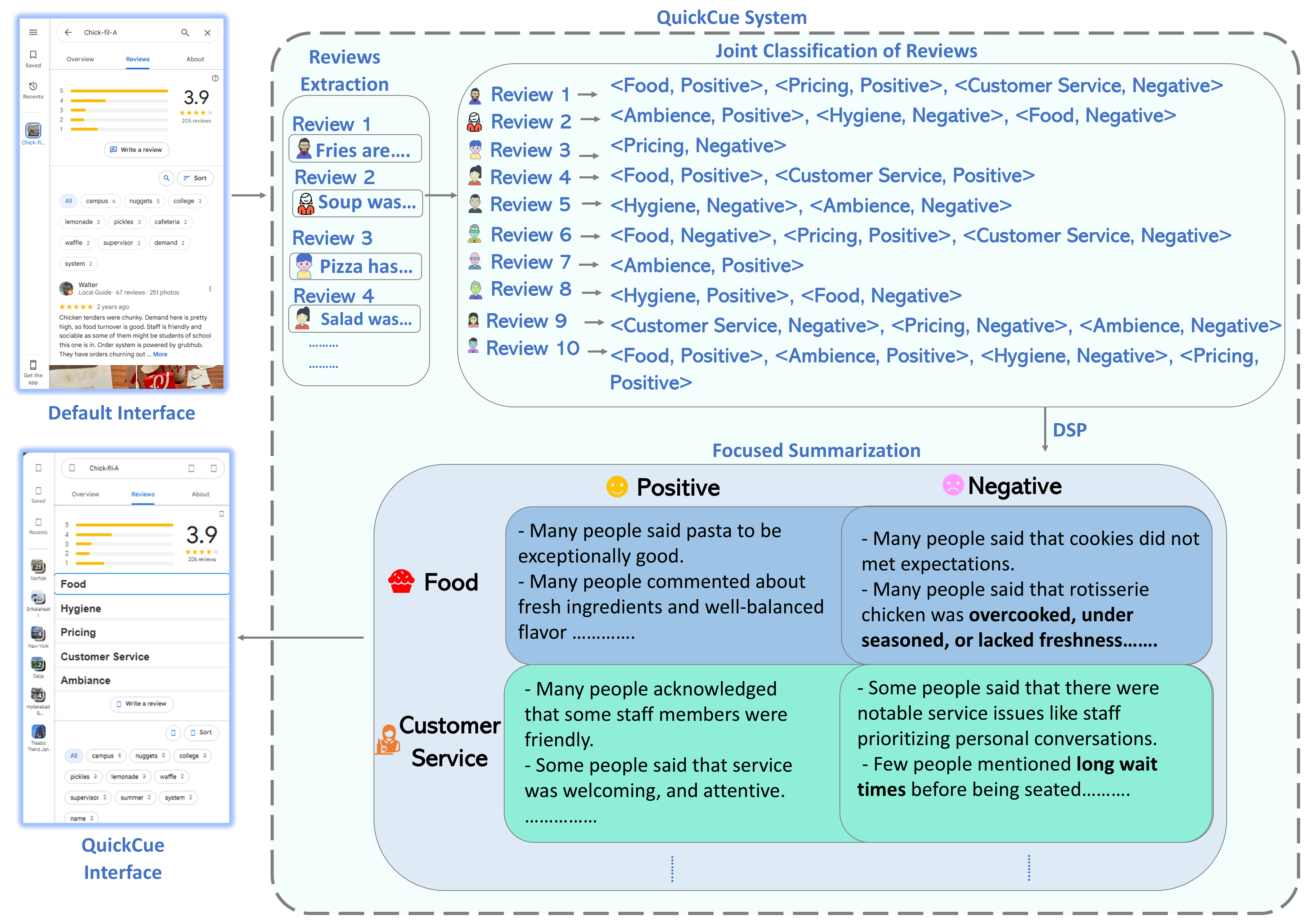}
    \caption{A workflow schematic depicting \sysname{}'s architecture.}
    \label{fig:architecture}
    \Description{A schematic illustrating QuickCue’s architecture, beginning on the left with an input interface that displays a restaurant’s star rating and user reviews. Beneath it lies the Virtual DOM layer. In the center, the QuickCue System box highlights “Joint Classification of Reviews,” showing how aspect-sentiment pairs are extracted (e.g., Service, Ambience, and Food). On the right, the “Focused Summarization” step presents a division into Positive and Negative summary categories, each containing summaries grouped by aspect, such as Food and Customer Service and their respective sentiment.}
\end{figure*}

Figure~\ref{fig:architecture} presents an operational schematic of the \sysname{} browser extension we built to thematically organize information in customer reviews and then present this processed information via a screen reader-friendly user interface (Figure~\ref{fig:intro}). For the case study, we chose the Google Maps platform, given that it is most popular platform for sharing reviews online, especially regarding restaurants~\cite{de2022reviewing}. As seen in the figure, \sysname{} augments the existing Google Maps user interface with additional accessible content in which the information in reviews are organized based on both their underlying aspect (e.g., food quality, pricing) and their sentiment (positive, negative). The first level comprises a list of five drop-down buttons, each corresponding to an aspect. The next level comprises positive and negative summaries pertaining to each of the five aspects. The last level simply lists the raw reviews classified as belonging to each of the aspect/sentiment pairs. This arrangement, designed based on the findings of the interview study, not only reduces information redundancy by enabling users to get a quick overview of the `good' and the `bad' of aspects the users care about, but also helps them focus on a subset of reviews pertaining to a specific aspect of interest, e.g., reviews that shed light on the negative experiences with customer service.   

To generate such an alternative presentation of information in reviews, \sysname{} addresses the following two main technical challenges, which primarily stem from the heterogeneous nature of customer reviews: (i) A review can contain information about multiple aspects; and (ii) A review can contain both positive and negative opinions about different aspects (e.g., \textit{food was good, but the table and seats were not properly cleaned}) and even within the same aspect (e.g., \textit{the staff were very friendly, but the wait was too long!}). To address these challenges, \sysname{} performs the following core operations by adapting the state-of-the-art large language models (LLMs): (i) Joint classification of reviews -- given a review, determine all the <aspect, sentiment> pairs that are applicable to that review based on its contained information; and (ii) Aspect-focused summarization of information in reviews -- given a set of reviews, generate a summary that only focuses on a particular aspect and sentiment. The details of these operations are provided in the next.

\subsection{Joint Classification of Reviews}

\sysname{} performs joint classification of reviews to group them according to the aspects and sentiments covered in their content. Specifically, for each review, \sysname{} determines all the <aspect,sentiment> pairs that are applicable to that review. For example, for the review "The food was delicious, but the service was slow.", \sysname{}'s joint classifier will output "[["Food," "Positive"],["Customer Service," "Negative"]]." As justified earlier, \sysname{} primarily looks for five aspects (food, ambiance, customer service, pricing, and hygiene) and two sentiments (positive and negative), while doing the classifications. 

To do the joint classification, \sysname{} leverages the GPT-4 large language model (LLM)~\cite{openai_gpt4}, due to its proven ability to reason over complex contexts and generate tailored outputs~\cite{thelwall2024chatgpt, hadar2023plasticity}. To instruct the LLM to accurately classify the reviews, we specifically adapted the Clue and Reasoning Prompting (CARP) strategy~\cite{sun-etal-2023-text}, given its suitability for this task. The CARP strategy enhances the classification performance by instructing the LLM to look for `clues' and use that in the reasoning process while determining the class of the input text. The clues may refer to a keyword, phrase, or contextual element extracted from the input text that provides evidence for classification. Since the original CARP prompting~\cite{sun-etal-2023-text} was intended for only sentiment classification, we modified it so as to make the LLM generate a set of aspect-sentiment pairs as output instead of a single classification label. The structure of our modified CARP prompt is shown below.

\begin{adjustwidth}{-0.3cm}{-0.3cm} 
\begin{quote}
\small
\begin{tcolorbox}[colframe=blue!70!black, colback=blue!5, sharp corners, boxrule=1pt]
\textcolor{black}{\textbf{Task Description:}} This is a joint aspect-sentiment classifier for restaurant reviews. \\

\textbf{First,} present \textcolor{red}{\textbf{CLUES}} (i.e., keywords, phrases, contextual information, semantic meaning, semantic relations, tones, references) that support the joint aspect-sentiment determination of input (look for clues related to \textit{Food, Ambiance, Customer Service, Pricing, Hygiene} for aspect, and clues related to \textit{positive, negative} for sentiment). \\

\textbf{Second,} deduce a diagnostic \textcolor{red}{\textbf{REASONING}} process from premises (i.e., clues, input) that supports the sentiment determination for each identified aspect. Note that an aspect can be identified multiple times in different locations of the input. \\

\textbf{Third,} determine the list of aspect-sentiment pairs present in the \textbf{INPUT}, considering the \textbf{CLUES} and the \textbf{REASONING} process. \\

Output all possible aspect-sentiment pairs after removing empty pairs if any. \\
For \textbf{ASPECT}, choose from the following predefined set of words: \textcolor{violet}{\texttt{[Food, Ambiance, Hygiene, Customer Service, Pricing]}}.\\ 
For \textbf{SENTIMENT}, choose from the following two words: \textcolor{violet}{\texttt{[Positive,Negative]}}\\

\textbf{EXAMPLES:} \\

\textbf{INPUT:} The ambiance was warm and inviting, but the pasta lacked seasoning and was undercooked.\\

\textbf{CLUES:} [ambiance, warm], [ambiance, inviting], [pasta, lacked seasoning], [pasta, undercooked] \\

\textbf{REASONING:} The terms ``warm'' and ``inviting'' suggest a welcoming and pleasant atmosphere, indicating a positive experience with ambiance. \\
The phrases ``lacked seasoning'' and ``undercooked'' indicate dissatisfaction with the food quality, suggesting a negative sentiment for food. \\

\textbf{ASPECT-SENTIMENT Pairs:} \textcolor{violet}{[``Ambiance'', ``Positive'']} \textcolor{violet}{[``Food'', ``Negative'']} \\

\textbf{INPUT:} \textcolor{black}{[Insert review text here]} \\
\end{tcolorbox}
\normalsize
\end{quote}
\end{adjustwidth}

\vspace{3pt}
As shown in the template, to further enhance the classification performance, we augmented the CARP prompt with few-shot examples. Based on the recommendation of prior work which advocated a minimum of 16 few-shot examples for the CARP prompt~\cite{sun-etal-2023-text}, we created 20 examples, i.e., 2 examples for each of the 10 aspect-sentiment pairs. We chose a diverse set of reviews for these examples, covering different restaurant cuisines and locations. 

After classifying each of the reviews using the joint classifier, \sysname{} categorizes the reviews based on the aspect-sentiment pairs, i.e., for each of the 10 aspect-sentiment pairs, \sysname{} identifies the corresponding subset of matching reviews, based on the classifier output. Note that, given the heterogeneity of reviews, a review can possibly be included in multiple subsets corresponding to different aspect-sentiment pairs. Each subset then serves as the input for generating \textit{focused} review summaries, as explained later in Section~\ref{sec:summarization}.

\noindent\textbf{Evaluation.} To evaluate the performance of the joint classifier, we created a ground truth dataset by manually annotating 50 reviews or examples which were randomly sampled across diverse restaurant cuisines and locations. The frequency breakdown for the 10 aspect-sentiment pairs in this dataset was as follows: [Food, Negative]: $24$, [Food, Positive]: $27$, [Customer Service, Negative]: $10$, [Customer Service, Positive]: $15$, [Pricing, Negative]: $9$, [Pricing, Positive]: $8$, [Ambiance, Negative]: $5$, [Ambiance, Positive]: $12$, [Hygiene, Negative]: $6$, [Hygiene, Positive]: $5$. The annotations were done manually by a research assistant and subsequently verified by two other research assistants. 

We experimented with two prompt variations: zero-shot (no illustrative examples) and few-shot (20 demonstrative examples), and we used the standard precision, recall, and F-1 metrics to measure performance. The few-shot method yielded a significantly better performance of $0.8001$ precision, $0.8201$ recall, and $0.8099$ F1-score (all averages) compared to the zero-shot method, which exhibited a much lower performance of $0.615$ precision, $0.625$ recall, and $0.6199$ F1-score. Note that the reported values are averages, since each review was tagged with multiple aspect-sentiment pairs; precision, recall, and F1-scores were computed for each review in the dataset and then averaged across all the reviews. Upon careful analysis of the classifier outputs, the majority of classification errors occurred when the the reviews had the same aspect occurring twice, once for each sentiment (e.g., \textit{pasta was excellent but the chicken tenders were cold}), thereby indicating that further experimentation with prompts is needed to differentiate between the two occurrences of the same aspect.

\subsection{Focused Summarization}\label{sec:summarization}
Once the subsets are identified for each aspect-sentiment pair, \sysname{} next generates \textit{focused} summaries pertaining to each pair. Recall that we need focused summarization due to the heterogeneous nature of information in reviews; general summarization techniques may omit salient information pertaining to the target aspect-sentiment pair and instead include information about other pairs in the generated summary. For this task, \sysname{} employs the Directional Stimulus Prompting (DSP) technique for LLMs~\cite{li2024guiding}. This prompting technique involves providing keywords as directional stimuli to tailor the summarization process via selective information prioritization. A snippet of our DSP prompt template is shown below.

\begin{adjustwidth}{-0.3cm}{-0.3cm} 
\begin{quote}
\small
\begin{tcolorbox}[colframe=blue!70!black, colback=blue!5, sharp corners, boxrule=1pt]
\textcolor{black}{\textbf{Task Instructions:}} Summarize the given reviews by focusing only on the specified main aspect and desired sentiment. Use the \textcolor{red}{\textbf{Directional Stimuli (keywords)}} for guidance. Ensure the generated summary \textbf{excludes unrelated aspects, redundant phrases, and undesired sentiments}, while keeping it concise and clear. \\

\textbf{Reviews:} \textcolor{black}{[Input reviews here]} 

\textcolor{red}{\textbf{Directional Stimuli:}} \\
\textbf{Main Aspect:} \textcolor{violet}{[Insert Desired Aspect]} \\
\textbf{Desired Sentiment:} \textcolor{violet}{[Insert Desired Sentiment]} \\

\textbf{Output Instruction:} \\
\textit{Generate the summary as a sequence of bullet points, with each point highlighting one salient feature uncovered about the specified aspect and desired sentiment.}\\

\textbf{Examples:} 

\textbf{Reviews:} {[Input reviews here]} \\
\textcolor{red}{\textbf{Directional Stimuli:}} \\
\textbf{Main Topic:} \textcolor{violet}{Customer Service} \\
\textbf{Sentiment:} \textcolor{violet}{Negative} \\
\textbf{Output Summary:} 
\begin{itemize}
    \renewcommand\labelitemi{-}
    \item Many customers complained about slow service, stating that their orders took significantly longer than expected.
    \item Some reviewers mentioned that the staff appeared inattentive and unresponsive, making it difficult to get assistance.
    \item Several customers reported that the staff lacked knowledge about the menu, leading to confusion when placing orders.
    \item Several reviews pointed out that employees lacked professionalism, often engaging in personal conversations rather than attending to customers.
    \item Many visitors expressed frustration over the lack of courtesy from staff, mentioning that employees were often rude or dismissive.
\end{itemize}

...\\

\textbf{INPUT:} \textcolor{black}{[Input reviews here]} \\

\end{tcolorbox}
\normalsize
\end{quote}
\end{adjustwidth}

As seen in the above prompt, we included few-shot examples~\cite{ahmed2022few} within the prompt template to improve the quality of the summarization. We specifically crafted 10 few-shot examples (one for each aspect-sentiment pair), each of which included a set of randomly sampled reviews and the corresponding handcrafted summary.

\vspace{3pt}
\noindent\textbf{Evaluation.} To evaluate the efficacy of our DSP prompt in generating focused summaries, we built a test dataset comprising 50 examples -- $5$ for each aspect-sentiment pair. The ground truth summaries for these examples were carefully handcrafted and verified to ensure that they contained all the salient pieces of information pertaining to the corresponding aspect-sentiment pairs. To evaluate the quality of generated summaries, we relied on manual human evaluation, given the proven unreliability of automatic evaluation methods~\cite{stent2005evaluating}. Specifically, 10 human annotators provided the following two metrics for each generated summary: (i) Factuality (1 for least accurate and 10 for highly accurate), which determines if all facts in the summary aligned with the reviews, and (ii) Noisiness (1 for highly noisy and 10 for least noisy), which evaluated the extent to which extraneous, off-topic information is present in a summary considering the target aspect-sentiment pair. For both of these metrics, the average score for each example (across the 10 annotators) was first computed, and then the overall average (i.e., average of averages) was computed across all the 50 examples in the test dataset. 

The average factuality and noisiness scores for the few-shot DSP prompt were $7.9$ and $8.3$ respectively. Annotators noted that reduced scores for factuality often stemmed from incomplete information in generated summaries. A closer inspection of low-scoring examples revealed that this was mostly due to conflicting and vague information in reviews. For example, for a particular dish, some reviews expressed positive sentiment due to its spiciness whereas a few other reviews expressed a positive sentiment indicating that the same dish was not that spicy. The generated summary however only included the latter aspect of the dish as one of the salient points. This highlights the inherent ambiguity due to the subjectivity of the peoples' perceptions regarding the different aspects. Note that \sysname{} follows a modular architecture, allowing the summarization module to be easily replaced with an improved version in future research.

\subsection{User Interface}
As mentioned earlier, \sysname{} inserts its content into the existing Google Maps as an augmentation, so that the user does not have to shift focus to another page (see Figure~\ref{fig:architecture}). As shown in Figure~\ref{fig:intro}, the content generated by \sysname{} is arranged in a hierarchy comprising three layers (i.e., aspects, focused summaries, and original reviews). Specifically, \sysname{} renders this hierarchy in HTML as an accordion, and automatically injects ARIA (Accessible Rich Internet Applications) attributes~\cite{w3c-aria} to make it accessible. Moreover, \sysname{} also adds tab-index attributes to relevant nodes in the accordion for enabling screen-reader users to easily navigate content at each layer using TAB and SHIFT+TAB hotkeys. \sysname{} allows the users to navigate down the hierarchy via the ENTER key, and navigate up the hierarchy using the ESCAPE key. In sum, \sysname{} simplifies interaction with information in reviews, limiting it to a few basic screen reader hotkeys.

\subsection{Additional Implementation Details}
The \sysname{} was implemented as a Google Chrome browser extension, following open-source guidelines for browser extensions~\cite{chrome_extensions}. Extraction of reviews from Google Maps was done by leveraging pre-defined XPath information identifying the HTML DOM nodes corresponding to these reviews. The extracted data was structured into JSON objects and transferred to the backend through RESTful API calls~\cite{flask_restful} for further processing.
Text preprocessing was performed using the NLTK~\cite{nltk} and spaCy~\cite{spacy} libraries, and noise elements such as emojis, out-of-vocabulary words, and excessive whitespace were filtered out using regular expressions~\cite{python_re}. All data exchanges between \sysname{} modules were in JSON for consistency and convenience. Integration of LLM into \sysname{} was done using the well-known LangChain framework~\cite{LangChain}.

%% file: Sections/5.Evaluation.tex
\section{User Study}
We conducted an IRB-approved user study with screen reader users to evaluate the usability of \sysname{} and compare it against the status quo. A total of $10$ blind participants\footnote{This is the typical sample size for research in this area, due to the difficulty in recruiting participants belonging to this community~\cite{bigham2007webinsitu, prakash2024all}.} ($4$ female, $6$ male) were recruited through email lists and snowball sampling. The participants had an average age of $31.4$ years (Median = $31.5$, Range = $22$–$43$). The inclusion criteria were: (i) familiarity with web screen reading and online review platforms; (ii) experience using the Chrome web browser and proficiency in JAWS; (iii) proficiency in using the standard QWERTY keyboard; and (iv) proficiency in communicating in English. To preserve external validity, we ensured that there was no overlap between the participant groups of this study and the prior interview study. All participants reported that they regularly engaged with customer reviews across various platforms, including shopping websites and food ordering services. The participant demographics are detailed in Table~\ref{table:participants}.

\begin{table}[t!]
\centering
{\def\arraystretch{1.1}
{\small
\begin{tabular}{ m{0.3cm} m{0.9cm} m{1.3cm} m{1.4cm} m{0.9cm}  m{1.5cm}}
  \toprule
  \multirow{2}{*}{\textbf{ID}} & 
  \multirow{2}{*}{\shortstack[l]{\textbf{Age/}\\ \textbf{Gender}}} & 
  \multirow{2}{*}{\shortstack[l]{\textbf{Age of}\\ \textbf{Vision Loss}}} & 
  \multirow{2}{*}{\shortstack[l]{\textbf{Occupation}}} & 
  \multirow{2}{*}{\shortstack[l]{\textbf{Screen}\\ \textbf{Reader}}} & 
  \multirow{2}{*}{\shortstack[l]{\textbf{Web}\\ \textbf{Experience}}} \\
  & & & & & \\
  \midrule
  P1  & 43/M & Since birth  & Teacher        & JAWS   & 7 years   \\ \hline
  P2  & 36/M & Age 8        & Unemployed     & JAWS   & 4 years   \\ \hline
  P3  & 28/M & Since birth  & Student        & NVDA   & 11 years   \\ \hline
  P4  & 23/M & Age 10       & Student        & JAWS   & 8 years    \\ \hline
  P5  & 36/F & Since birth  & Social Worker  & JAWS   & 5 years   \\ \hline
  P6  & 32/M & Age 6        & Teacher        & JAWS   & 4 years   \\ \hline
  P7  & 25/F & NA & Unemployed  & NVDA   & 2 years   \\ \hline
  P8  & 38/M & NA & Social Worker & JAWS   & 6 years   \\ \hline
  P9  & 31/F & Since birth  & Teacher        & JAWS   & 5 years   \\ \hline
  P10 & 22/F & Age 8        & Student        & JAWS   & 6 years    \\
  \bottomrule
\end{tabular}
}
}
\Description[Participant demographics]{All information was self-reported by the participants.}
\caption{Participant demographics. All information was self-reported by the participants.}
\label{table:participants}
\end{table}

\subsection{Design}
In a within-subject experimental design, each participant was asked to freely explore and compare two restaurants on Google Maps based on their reviews, under the following two conditions:
\vspace{0.2cm}
\begin{itemize}
    \item \textbf{Screen Reader}: The status quo baseline, where the participants used their screen reader to peruse reviews in the default Google Maps user interface.
    \item \textbf{\sysname{}}: The participants used their screen reader to interact with the augmented Google Maps user interface, containing the accordion generated by our \sysname{} assistive tool.
\end{itemize}

Influenced by the insights from the interview study, we chose a free-form comparison task to emulate real-world interaction scenarios in which users typically navigate reviews of multiple restaurants before making their decisions. To mitigate learning effects and avoid confounding variables, we selected four different restaurants for the two tasks. Additionally, we ensured that \sysname{} accurately retrieved all reviews to prevent any confounding effects of retrieval accuracy. The assignment of restaurants to conditions and the ordering of conditions were counterbalanced across study participants using the well-known Latin Square method~\cite{{bradley1958complete}}. A maximum of $30$ minutes was allotted for each task.

\subsection{Procedure}
At the beginning of the study, the experimenter explained the study objectives to the participant, and obtained an informed consent. This was followed by a practice session where the participant was allowed to familiarize with the \sysname{} interface, refresh memories regarding screen reader hotkeys, and making any adjustments to the screen reader configuration (e.g., adjust speech rate). Note that all participants did the tasks on a Windows ThinkPad laptop provided by the experimenter, with all the necessary software installed, and also connected to an external standard QWERTY keyboard familiar to all the participants. The experimenter then asked the participant to complete the tasks in the pre-determined counterbalanced order. After the tasks, the experimenter administered the standard questionnaires, namely the System Usability Scale (SUS)~\cite{brooke1996sus} to assess usability, and the NASA Task Load Index (NASA-TLX)~\cite{hart1998development} to evaluate perceived workload. Lastly, the experimenter debriefed the participant in an exit interview, encouraging to provide subjective feedback, including the experience with \sysname{}, difficulties while doing the tasks, and suggestions for improvement. All interactions were conducted in English. To prioritize participants' well-being, they were informed that they could take breaks or withdraw from the study at any time. The participants received \$30 as compensation for their time.

\subsection{Data Collection and Analysis}
The experimenter documented participants' responses to the SUS and NASA-TLX questionnaires, recorded their think-aloud utterances during task execution, and observed their screen reader interaction behavior throughout the study. The SUS and NASA-TLX responses were analyzed using descriptive and inferential statistical methods. For qualitative data, we applied grounded theory methods~\cite{oktay2012grounded}, specifically the open coding and axial coding techniques~\cite{saldana2021coding} to systematically iterate over transcribed participant responses and uncover recurring themes and key insights. We present our findings next.

\subsection{Results}
\begin{figure}[t!]   
    \centering 
    \includegraphics[width=\columnwidth]{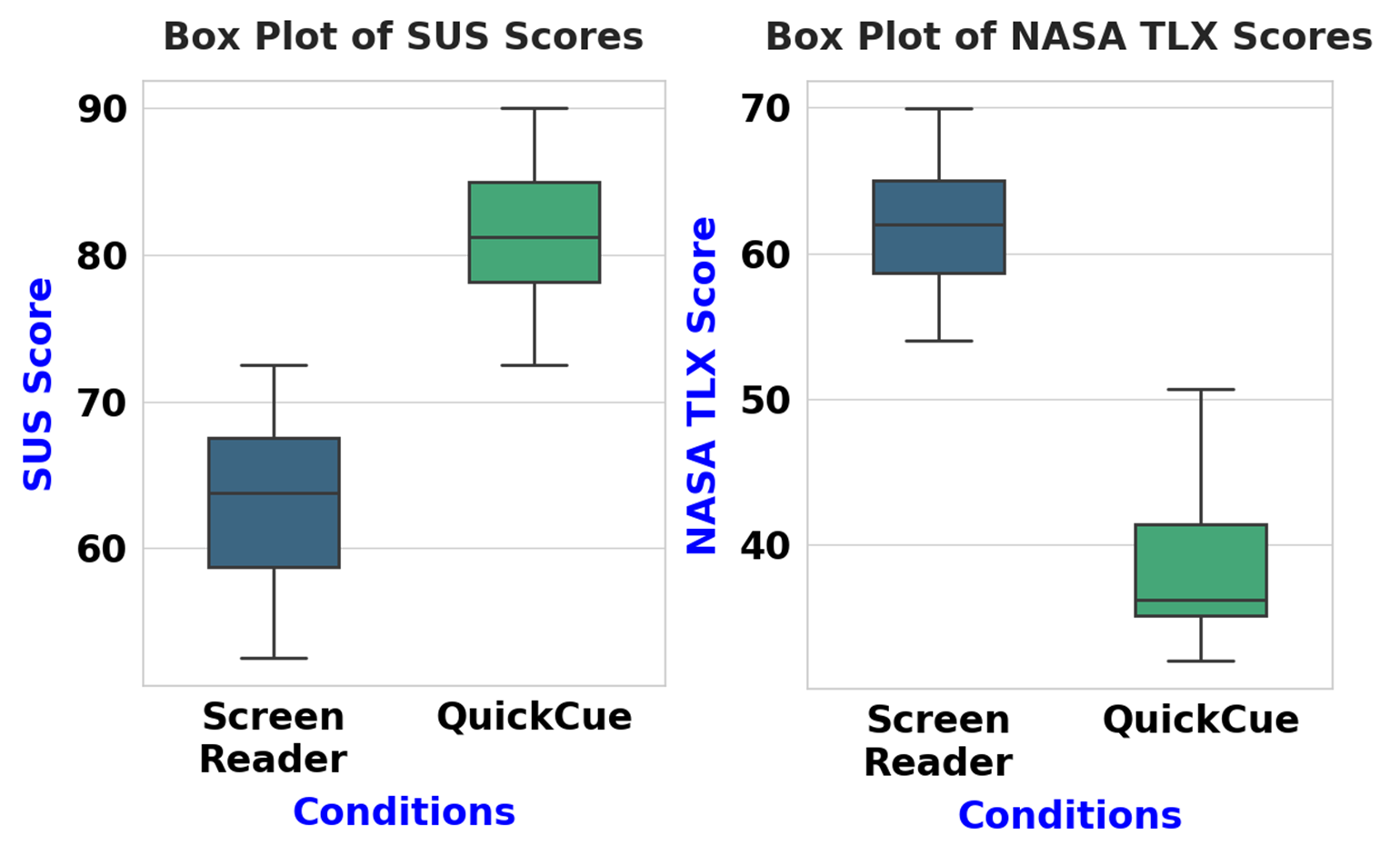}
    \caption{(Left) SUS scores, (right) NASA-TLX scores.}
    \label{fig:nasatlx}
    \Description{Figure shows two side-by-side box plots comparing the Screen Reader and QuickCue conditions. The left plot, labeled “Box Plot of SUS Scores,” displays higher median SUS scores for QuickCue, while the right plot, labeled “Box Plot of NASA TLX Scores,” shows lower median NASA TLX scores for QuickCue, indicating improved usability and reduced workload relative to the Screen Reader condition.}
\end{figure}

\subsubsection{Usability}
As mentioned earlier, SUS questionnaire~\cite{brooke1996sus} was used to evaluate usability. Specifically, SUS asks the participants to rate ten statements on a 5-point Likert scale ranging from 1 to 5, with 1 indicating ``strongly disagree'' and 5 signifying ``strongly agree.'' A final SUS score between 0 and 100 is then calculated by assimilating the individual ratings based on a predefined formula, and higher scores indicate better usability. The SUS scores for the two experimental conditions are shown in Figure~\ref{fig:nasatlx}, where it can be clearly observed that the \sysname{} condition received significantly higher scores compared to the default screen reader condition. Specifically, the Screen Reader condition received an average SUS score of $63.25$ (median = $63.75$, min = $52.5$, max = $72.5$), whereas the  \sysname{} condition received an average SUS score of $81.5$ (median = $81.25$, min = $72.5$, max = $90.0$). A one-way ANOVA test confirmed that the difference in usability scores between the two conditions was statistically significant $F = 45.03$, $p = 2.72 \times 10^{-6}$.

\subsubsection{Task Workload}
We employed the standard NASA-TLX questionnaire~\cite{hart1998development} to assess task workload. Like SUS, NASA-TLX also assimilates user's ratings into a score between 0 and 100, however, lower TLX scores indicate less workload and therefore better performance and user experience. The TLX score statistics for the two conditions are shown in Figure~\ref{fig:nasatlx}, where it can be observed that \sysname{} significantly decreased the task workload, thereby substantially improving participants' user experience. Specifically, in the Screen Reader condition, the average TLX score was $62.09$ (median = $62.0$, min = $54.0$, max = $69.93$) and in the \sysname{} condition, the average TLX score was $38.37$ (median = $36.17$, min = $32.0$, max = $50.67$). This difference in average TLX scores between conditions was also found to be statistically significant (One-way ANOVA, $F = 99.27$, $P = 9.45 \times 10^{-9}$). A closer inspection of individual ratings revealed that the \textit{Effort} and \textit{Temporal Demand} TLX sub-scales contributed relatively more to the significant difference between conditions, compared to the other four sub-scales.  
\subsubsection{Qualitative Feedback.}
The following core insights were uncovered from the qualitative analysis of the participants' subjective feedback during the exit interviews. \\

\noindent \textbf{Simplistic design and ease of use. }
A majority (7) of participants attributed their higher usability perception of \sysname{} to its simplistic design, which allowed them to navigate it using simple keyboard shortcuts that they were already familiar with. Additionally, they expressed appreciation for the \textit{``summary feature''}, stating that it helped them to \textit{``listen less and learn more''} and \textit{``listen to only what they wanted''} about a restaurant. Regarding this, P3 said, `` If I have these summaries, I will not at all listen to the reviews. It is extremely frustrating to listen to a lot of irrelevant and repeated feedback that barely tells me anything about what I would like to know about the food and the experience.'' The experimenter also noted such an interaction behavior during the study, where some of the participants did not bother going through the original reviews, and instead just listened to the summaries pertaining to a few aspect-sentiment pairs of interest. Five participants further stated that \sysname{} would enable them to ``explore more'' and try ordering new dishes instead of the ordering the ``same old tried-and-tested'' items they are already familiar with at a given restaurant. For instance, P2 mentioned, \textit{``This is very helpful...you know...I don't cook much, but when I order, it's always the same food because it takes a lot of time to sit and read reviews, and find something else...especially after a long day's work. If there's an easier way, of course, I'd use it to know what else is good.''}

\vspace{3pt}
\noindent \textbf{Extension of \sysname{} to other review platforms. } 
All participants appreciated how perusing user reviews on Google Maps was \textit{``quite organized''} and \textit{``less boring''} with \sysname{}, enabling them to explore others' opinions with greater interest and helping them feel \textit{``more confident''} in their dining decisions. Seven participants inquired if the system could be extended to other platforms, particularly e-commerce, where they prefer to read user reviews before making a purchase. For instance, P8 asked, \textit{``Does this work on Amazon? I shop quite a lot there, and this would definitely help me make better purchase decisions. I'd like to know what others are saying about a product before I buy it...like its quality, whether it's worth the price, or if there are any updates or improvements.''}

\vspace{3pt}
\noindent \textbf{Repetitive search hinders restaurant comparison. } 
Nearly all participants (9) mentioned that during the comparison task, they had to repeat the tedious process of searching for pieces of information pertaining to a specific aspect of interest, when then navigated to the reviews of the second restaurant from the reviews of the first restaurant. The experimenter also noted some of the think-aloud utterances that corroborate this statement, e.g., \textit{``Okay, now I have to do it again. Let's find out where I can find comments about the taste and price of burgers.''} Five participants further explained that such a repetitive search process was simply ``too tiring'' without additional support, and that \sysname{} helped them reduce this effort to a ``large extent''. Nonetheless, this feedback highlights the need for personalized summarization to enable users to prioritize their favorite aspects across multiple restaurants, thereby facilitating convenient comparisons between restaurants.

%% file: Sections/6.Discussion.tex
\section{Discussion}

\subsection{Limitations}

One limitation of our work is that we evaluated \sysname{} exclusively with JAWS screen reader users. Although JAWS is the most popular screen reader, many blind users also use other screen readers such as NVDA and VoiceOver \cite{webaim2019screen}. While  \sysname{} conceptually is screen reader-agnostic and will therefore likely produce similar results in the case of other screen reader users, the evaluation must nonetheless be conducted to validate our hypotheses. 

Another limitation of our prototype is that it currently supports only desktop and laptop platforms. Research highlights a growing trend of users relying on smartphones to read and interact with online reviews, necessitating the adaptation of \sysname{} for mobile platforms. However, this transition poses challenges, as mobile web browsers generally lack support for extensions. To overcome this limitation, we plan to explore alternative solutions, such as developing standalone service applications, to extend \sysname{}'s functionality to smartphone users.

A third limitation is that our few-shot learning and testing examples for joint classification and focused summarization were confined to English-language restaurant reviews, leaving the algorithm's effectiveness in other languages unexplored. The sizes of the samples too were relatively small due to the large amount of manual effort involved in building these test datasets. Moreover, \sysname{} is currently restricted to the Chrome web browser. Although Chrome is the most commonly used browser among blind users, a significant portion of users still rely on alternatives such as Firefox~\cite{webaim2015screen} and Safari. Future work will therefore also focus on expanding \sysname{}'s compatibility to other languages and additional web browsers.

Another limitation of our work was that the technique to extract the reviews from the Google Maps webpage consisted of handcrafted rules based on predefined XPath patterns. For enabling \sysname{} to function end-to-end, algorithms need to be devised to automatically detect and extract user reviews from webpages.

Furthermore, our evaluation was mostly qualitative, lacking quantitative metrics such as task completion time or error rates. While this approach provided valuable user experience insights, in a future study with a larger group of participants, we plan to design tasks that will facilitate quantitative analysis and therefore a more comprehensive assessment of \sysname{}'s effectiveness.

Also, as mentioned earlier, \sysname{} was specifically designed and tested only for the review sections of restaurant menus, and its effectiveness on other types of websites is yet to be explored. However, given the modular and generalizable architecture of \sysname{}, future research could focus on expanding its functionality to other website genres such as e-commerce platforms, classifieds, and entertainment sites, to evaluate its broader applicability. 

\subsection{Personalized User Preferences}
In the subjective feedback, many participants highlighted the need for personalization in \sysname{}, specifically the ability to store user preferences such as favorite aspect-sentiment pairs, and ensuring it automatically applies across all restaurant menus. Based on this feedback, we plan to extend \sysname{} to support user-driven customization~\cite{bila2007pagetailor,nebeling2013crowdadapt}. To implement this, we will develop a pop-up interface that allows users to pre-select their preferences from predefined aspect-sentiment pair options. These selections will then be processed through a custom filtering algorithm, dynamically generating a user interface that reflects the chosen settings. In future work, we aim to fully integrate personalized support within \sysname{}, enhancing usability and delivering a more tailored, user-centric experience.

\subsection{Generalization Beyond Restaurant Reviews}
Online customer reviews provide valuable, topical, and relevant feedback on service features and user experiences~\cite{valdivia2017sentiment}. Numerous studies have investigated customer reviews in the restaurant sector, where the experiential nature of dining amplifies the impact of reviews and user comments~\cite{li2023automating}. Given this, our paper focused on developing \sysname{} using restaurant reviews as a case study to enhance the reviews' usability and improve the overall user experience for blind users. However, \sysname{} is not strictly limited to restaurant reviews, as its modular architecture was designed for inherent scalability across different domains. It consists of two core components: joint classification and focused summarization, enabling seamless adaptation to various review-based platforms, such as product reviews on e-commerce sites or classified advertisements. This flexibility requires minimal re-engineering, primarily involving data-driven modifications. For instance, in this study, we classify reviews based on five predefined aspects; transitioning to a different domain would require identifying and extracting relevant domain-specific aspects. Moreover, the prompt templates utilized in our approach are specifically curated for restaurant reviews and would also need to be adapted to align with the contextual requirements of other domains.

%% file: Sections/7.Conclusion.tex
\section{Conclusion}
The current layout of customer reviews regarding restaurants primarily caters to the preferences and ease of sighted users. For blind users, however, this arrangement results in a tedious and frustrating content-consumption experience, requiring them to traverse large volumes of text while often encountering irrelevant content. To address this usability gap, we developed \sysname{}, an intelligent assistive tool embodied as a browser extension specifically designed for blind screen reader users to conveniently access information in online customer reviews, specifically those pertaining to Google Maps. \sysname{} streamlines access to review sections, allowing users to efficiently search for relevant information and compare restaurant menus more effectively, thereby enhancing decision-making. The \sysname{} organizes the review section by breaking it down into selectable aspects (e.g., food, service, ambiance), followed by the presentation of positive and negative summaries for each aspect to provide a quick overview. In a user study with $10$ blind participants, \sysname{} significantly outperformed the status quo regarding usability and overall user experience.

%% file: Main.bbl

\begin{thebibliography}{93}


\ifx \showCODEN    \undefined \def \showCODEN     #1{\unskip}     \fi
\ifx \showDOI      \undefined \def \showDOI       #1{#1}\fi
\ifx \showISBNx    \undefined \def \showISBNx     #1{\unskip}     \fi
\ifx \showISBNxiii \undefined \def \showISBNxiii  #1{\unskip}     \fi
\ifx \showISSN     \undefined \def \showISSN      #1{\unskip}     \fi
\ifx \showLCCN     \undefined \def \showLCCN      #1{\unskip}     \fi
\ifx \shownote     \undefined \def \shownote      #1{#1}          \fi
\ifx \showarticletitle \undefined \def \showarticletitle #1{#1}   \fi
\ifx \showURL      \undefined \def \showURL       {\relax}        \fi
\providecommand\bibfield[2]{#2}
\providecommand\bibinfo[2]{#2}
\providecommand\natexlab[1]{#1}
\providecommand\showeprint[2][]{arXiv:#2}

\bibitem[Abascal et~al\mbox{.}(2019)]%
        {abascal2019tools}
\bibfield{author}{\bibinfo{person}{Julio Abascal}, \bibinfo{person}{Myriam Arrue}, {and} \bibinfo{person}{Xabier Valencia}.} \bibinfo{year}{2019}\natexlab{}.
\newblock \showarticletitle{Tools for web accessibility evaluation}.
\newblock \bibinfo{journal}{\emph{Web accessibility: a foundation for research}} (\bibinfo{year}{2019}), \bibinfo{pages}{479--503}.
\newblock


\bibitem[Abou-Zahra(2008)]%
        {abou2008web}
\bibfield{author}{\bibinfo{person}{Shadi Abou-Zahra}.} \bibinfo{year}{2008}\natexlab{}.
\newblock \showarticletitle{Web accessibility evaluation}.
\newblock \bibinfo{journal}{\emph{Web accessibility: A foundation for research}} (\bibinfo{year}{2008}), \bibinfo{pages}{79--106}.
\newblock


\bibitem[Abuaddous et~al\mbox{.}(2016)]%
        {abuaddous2016web}
\bibfield{author}{\bibinfo{person}{Hayfa~Y Abuaddous}, \bibinfo{person}{Mohd~Zalisham Jali}, {and} \bibinfo{person}{Nurlida Basir}.} \bibinfo{year}{2016}\natexlab{}.
\newblock \showarticletitle{Web accessibility challenges}.
\newblock \bibinfo{journal}{\emph{International Journal of Advanced Computer Science and Applications}} \bibinfo{volume}{7}, \bibinfo{number}{10} (\bibinfo{year}{2016}).
\newblock


\bibitem[Achiam et~al\mbox{.}(2023)]%
        {achiam2023gpt}
\bibfield{author}{\bibinfo{person}{Josh Achiam}, \bibinfo{person}{Steven Adler}, \bibinfo{person}{Sandhini Agarwal}, \bibinfo{person}{Lama Ahmad}, \bibinfo{person}{Ilge Akkaya}, \bibinfo{person}{Florencia~Leoni Aleman}, \bibinfo{person}{Diogo Almeida}, \bibinfo{person}{Janko Altenschmidt}, \bibinfo{person}{Sam Altman}, \bibinfo{person}{Shyamal Anadkat}, {et~al\mbox{.}}} \bibinfo{year}{2023}\natexlab{}.
\newblock \showarticletitle{Gpt-4 technical report}.
\newblock \bibinfo{journal}{\emph{arXiv preprint arXiv:2303.08774}} (\bibinfo{year}{2023}).
\newblock


\bibitem[Ahmed and Devanbu(2022)]%
        {ahmed2022few}
\bibfield{author}{\bibinfo{person}{Toufique Ahmed} {and} \bibinfo{person}{Premkumar Devanbu}.} \bibinfo{year}{2022}\natexlab{}.
\newblock \showarticletitle{Few-shot training llms for project-specific code-summarization}. In \bibinfo{booktitle}{\emph{Proceedings of the 37th IEEE/ACM international conference on automated software engineering}}. \bibinfo{pages}{1--5}.
\newblock


\bibitem[AI(2024)]%
        {spacy}
\bibfield{author}{\bibinfo{person}{Explosion AI}.} \bibinfo{year}{2024}\natexlab{}.
\newblock \bibinfo{title}{spaCy: Industrial-Strength Natural Language Processing}.
\newblock
\newblock
\urldef\tempurl%
\url{https://spacy.io/}
\showURL{%
\tempurl}
\newblock
\shownote{Accessed: March 3, 2025}.


\bibitem[{Apple Inc.}(2022)]%
        {voiceover}
\bibfield{author}{\bibinfo{person}{{Apple Inc.}}} \bibinfo{year}{2022}\natexlab{}.
\newblock \bibinfo{title}{Accessibility - Vision - Apple}.
\newblock \bibinfo{howpublished}{\url{https://www.apple.com/accessibility/vision/}}.
\newblock


\bibitem[Ara et~al\mbox{.}(2020)]%
        {ara2020understanding}
\bibfield{author}{\bibinfo{person}{Jinat Ara}, \bibinfo{person}{Md~Toufique Hasan}, \bibinfo{person}{Abdullah Al~Omar}, {and} \bibinfo{person}{Hanif Bhuiyan}.} \bibinfo{year}{2020}\natexlab{}.
\newblock \showarticletitle{Understanding customer sentiment: Lexical analysis of restaurant reviews}. In \bibinfo{booktitle}{\emph{2020 IEEE Region 10 Symposium (TENSYMP)}}. IEEE, \bibinfo{pages}{295--299}.
\newblock


\bibitem[Ashok et~al\mbox{.}(2019)]%
        {ashok2019auto}
\bibfield{author}{\bibinfo{person}{Vikas Ashok}, \bibinfo{person}{Syed~Masum Billah}, \bibinfo{person}{Yevgen Borodin}, {and} \bibinfo{person}{IV Ramakrishnan}.} \bibinfo{year}{2019}\natexlab{}.
\newblock \showarticletitle{Auto-suggesting browsing actions for personalized web screen reading}. In \bibinfo{booktitle}{\emph{Proceedings of the 27th ACM Conference on User Modeling, Adaptation and Personalization}}. \bibinfo{pages}{252--260}.
\newblock


\bibitem[Ashok et~al\mbox{.}(2014)]%
        {ashok2014wizard}
\bibfield{author}{\bibinfo{person}{Vikas Ashok}, \bibinfo{person}{Yevgen Borodin}, \bibinfo{person}{Svetlana Stoyanchev}, \bibinfo{person}{Yuri Puzis}, {and} \bibinfo{person}{IV Ramakrishnan}.} \bibinfo{year}{2014}\natexlab{}.
\newblock \showarticletitle{Wizard-of-Oz evaluation of speech-driven web browsing interface for people with vision impairments}. In \bibinfo{booktitle}{\emph{Proceedings of the 11th Web for All Conference}}. \bibinfo{pages}{1--9}.
\newblock


\bibitem[Ashok et~al\mbox{.}(2017)]%
        {ashok2017web}
\bibfield{author}{\bibinfo{person}{Vikas Ashok}, \bibinfo{person}{Yury Puzis}, \bibinfo{person}{Yevgen Borodin}, {and} \bibinfo{person}{IV Ramakrishnan}.} \bibinfo{year}{2017}\natexlab{}.
\newblock \showarticletitle{Web screen reading automation assistance using semantic abstraction}. In \bibinfo{booktitle}{\emph{Proceedings of the 22nd International Conference on Intelligent User Interfaces}}. \bibinfo{pages}{407--418}.
\newblock


\bibitem[Aydin et~al\mbox{.}(2020)]%
        {aydin2020sail}
\bibfield{author}{\bibinfo{person}{Ali~Selman Aydin}, \bibinfo{person}{Shirin Feiz}, \bibinfo{person}{Vikas Ashok}, {and} \bibinfo{person}{IV Ramakrishnan}.} \bibinfo{year}{2020}\natexlab{}.
\newblock \showarticletitle{Sail: Saliency-driven injection of aria landmarks}. In \bibinfo{booktitle}{\emph{Proceedings of the 25th international conference on intelligent user interfaces}}. \bibinfo{pages}{111--115}.
\newblock


\bibitem[Bigham et~al\mbox{.}(2007)]%
        {bigham2007webinsitu}
\bibfield{author}{\bibinfo{person}{Jeffrey~P Bigham}, \bibinfo{person}{Anna~C Cavender}, \bibinfo{person}{Jeremy~T Brudvik}, \bibinfo{person}{Jacob~O Wobbrock}, {and} \bibinfo{person}{Richard~E Ladner}.} \bibinfo{year}{2007}\natexlab{}.
\newblock \showarticletitle{WebinSitu: a comparative analysis of blind and sighted browsing behavior}. In \bibinfo{booktitle}{\emph{Proceedings of the 9th International ACM SIGACCESS Conference on Computers and Accessibility}}. \bibinfo{pages}{51--58}.
\newblock


\bibitem[Bila et~al\mbox{.}(2007)]%
        {bila2007pagetailor}
\bibfield{author}{\bibinfo{person}{Nilton Bila}, \bibinfo{person}{Troy Ronda}, \bibinfo{person}{Iqbal Mohomed}, \bibinfo{person}{Khai~N Truong}, {and} \bibinfo{person}{Eyal De~Lara}.} \bibinfo{year}{2007}\natexlab{}.
\newblock \showarticletitle{Pagetailor: reusable end-user customization for the mobile web}. In \bibinfo{booktitle}{\emph{Proceedings of the 5th international conference on Mobile systems, applications and services}}. \bibinfo{pages}{16--29}.
\newblock


\bibitem[Billah et~al\mbox{.}(2017)]%
        {billah2017speed}
\bibfield{author}{\bibinfo{person}{Syed~Masum Billah}, \bibinfo{person}{Vikas Ashok}, \bibinfo{person}{Donald~E Porter}, {and} \bibinfo{person}{IV Ramakrishnan}.} \bibinfo{year}{2017}\natexlab{}.
\newblock \showarticletitle{Speed-dial: A surrogate mouse for non-visual web browsing}. In \bibinfo{booktitle}{\emph{Proceedings of the 19th International ACM SIGACCESS Conference on Computers and Accessibility}}. \bibinfo{pages}{110--119}.
\newblock


\bibitem[Bird et~al\mbox{.}(2024)]%
        {nltk}
\bibfield{author}{\bibinfo{person}{Steven Bird}, \bibinfo{person}{Ewan Klein}, {and} \bibinfo{person}{Edward Loper}.} \bibinfo{year}{2024}\natexlab{}.
\newblock \bibinfo{title}{Natural Language Toolkit (NLTK)}.
\newblock
\newblock
\urldef\tempurl%
\url{https://www.nltk.org/}
\showURL{%
\tempurl}
\newblock
\shownote{Accessed: March 3, 2025}.


\bibitem[Bradley(1958)]%
        {bradley1958complete}
\bibfield{author}{\bibinfo{person}{James~V. Bradley}.} \bibinfo{year}{1958}\natexlab{}.
\newblock \showarticletitle{Complete Counterbalancing of Immediate Sequential Effects in a Latin Square Design}.
\newblock \bibinfo{journal}{\emph{J. Amer. Statist. Assoc.}} \bibinfo{volume}{53}, \bibinfo{number}{282} (\bibinfo{year}{1958}), \bibinfo{pages}{525--528}.
\newblock
\urldef\tempurl%
\url{https://doi.org/10.1080/01621459.1958.10501456}
\showDOI{\tempurl}
\showeprint{https://amstat.tandfonline.com/doi/pdf/10.1080/01621459.1958.10501456}


\bibitem[Brajnik(2004)]%
        {brajnik2004comparing}
\bibfield{author}{\bibinfo{person}{Giorgio Brajnik}.} \bibinfo{year}{2004}\natexlab{}.
\newblock \showarticletitle{Comparing accessibility evaluation tools: a method for tool effectiveness}.
\newblock \bibinfo{journal}{\emph{Universal access in the information society}}  \bibinfo{volume}{3} (\bibinfo{year}{2004}), \bibinfo{pages}{252--263}.
\newblock


\bibitem[Brock et~al\mbox{.}(2015)]%
        {brock2015interactivity}
\bibfield{author}{\bibinfo{person}{Anke~M Brock}, \bibinfo{person}{Philippe Truillet}, \bibinfo{person}{Bernard Oriola}, \bibinfo{person}{Delphine Picard}, {and} \bibinfo{person}{Christophe Jouffrais}.} \bibinfo{year}{2015}\natexlab{}.
\newblock \showarticletitle{Interactivity improves usability of geographic maps for visually impaired people}.
\newblock \bibinfo{journal}{\emph{Human--Computer Interaction}} \bibinfo{volume}{30}, \bibinfo{number}{2} (\bibinfo{year}{2015}), \bibinfo{pages}{156--194}.
\newblock


\bibitem[Brooke(1996)]%
        {brooke1996sus}
\bibfield{author}{\bibinfo{person}{John Brooke}.} \bibinfo{year}{1996}\natexlab{}.
\newblock \showarticletitle{Sus: a “quick and dirty’usability}.
\newblock \bibinfo{journal}{\emph{Usability evaluation in industry}} \bibinfo{volume}{189}, \bibinfo{number}{3} (\bibinfo{year}{1996}).
\newblock


\bibitem[Caldwell et~al\mbox{.}(2008)]%
        {caldwell2008web}
\bibfield{author}{\bibinfo{person}{Ben Caldwell}, \bibinfo{person}{Michael Cooper}, \bibinfo{person}{Loretta~Guarino Reid}, \bibinfo{person}{Gregg Vanderheiden}, \bibinfo{person}{Wendy Chisholm}, \bibinfo{person}{John Slatin}, {and} \bibinfo{person}{Jason White}.} \bibinfo{year}{2008}\natexlab{}.
\newblock \showarticletitle{Web content accessibility guidelines (WCAG) 2.0}.
\newblock \bibinfo{journal}{\emph{WWW Consortium (W3C)}}  \bibinfo{volume}{290} (\bibinfo{year}{2008}), \bibinfo{pages}{1--34}.
\newblock


\bibitem[Chase(2022)]%
        {LangChain}
\bibfield{author}{\bibinfo{person}{Harrison Chase}.} \bibinfo{year}{2022}\natexlab{}.
\newblock \bibinfo{title}{LangChain: Building Applications with LLMs through Composability}.
\newblock \bibinfo{howpublished}{GitHub Repository}.
\newblock
\urldef\tempurl%
\url{https://github.com/langchain-ai/langchain}
\showURL{%
\tempurl}


\bibitem[Chhikara et~al\mbox{.}(2024)]%
        {chhikara2024lamsum}
\bibfield{author}{\bibinfo{person}{Garima Chhikara}, \bibinfo{person}{Anurag Sharma}, \bibinfo{person}{V Gurucharan}, \bibinfo{person}{Kripabandhu Ghosh}, {and} \bibinfo{person}{Abhijnan Chakraborty}.} \bibinfo{year}{2024}\natexlab{}.
\newblock \showarticletitle{LaMSUM: Amplifying Voices Against Harassment through LLM Guided Extractive Summarization of User Incident Reports}.
\newblock \bibinfo{journal}{\emph{arXiv preprint arXiv:2406.15809}} (\bibinfo{year}{2024}).
\newblock


\bibitem[Chisholm et~al\mbox{.}(2001)]%
        {chisholm2001web}
\bibfield{author}{\bibinfo{person}{Wendy Chisholm}, \bibinfo{person}{Gregg Vanderheiden}, {and} \bibinfo{person}{Ian Jacobs}.} \bibinfo{year}{2001}\natexlab{}.
\newblock \showarticletitle{Web content accessibility guidelines 1.0}.
\newblock \bibinfo{journal}{\emph{Interactions}} \bibinfo{volume}{8}, \bibinfo{number}{4} (\bibinfo{year}{2001}), \bibinfo{pages}{35--54}.
\newblock


\bibitem[Consortium et~al\mbox{.}(1999)]%
        {world1999web}
\bibfield{author}{\bibinfo{person}{World Wide~Web Consortium} {et~al\mbox{.}}} \bibinfo{year}{1999}\natexlab{}.
\newblock \showarticletitle{Web content accessibility guidelines 1.0}.
\newblock  (\bibinfo{year}{1999}).
\newblock


\bibitem[Dash et~al\mbox{.}(2021)]%
        {dash2021personalized}
\bibfield{author}{\bibinfo{person}{Anupam Dash}, \bibinfo{person}{Dongsong Zhang}, {and} \bibinfo{person}{Lina Zhou}.} \bibinfo{year}{2021}\natexlab{}.
\newblock \showarticletitle{Personalized ranking of online reviews based on consumer preferences in product features}.
\newblock \bibinfo{journal}{\emph{International Journal of Electronic Commerce}} \bibinfo{volume}{25}, \bibinfo{number}{1} (\bibinfo{year}{2021}), \bibinfo{pages}{29--50}.
\newblock


\bibitem[De~Boeck et~al\mbox{.}(2022)]%
        {de2022reviewing}
\bibfield{author}{\bibinfo{person}{Kevin De~Boeck}, \bibinfo{person}{Jenno Verdonck}, \bibinfo{person}{Michiel Willocx}, \bibinfo{person}{Jorn Lapon}, {and} \bibinfo{person}{Vincent Naessens}.} \bibinfo{year}{2022}\natexlab{}.
\newblock \showarticletitle{Reviewing review platforms: a privacy perspective}. In \bibinfo{booktitle}{\emph{Proceedings of the 17th International Conference on Availability, Reliability and Security}}. \bibinfo{pages}{1--10}.
\newblock


\bibitem[Developers(2024a)]%
        {flask_restful}
\bibfield{author}{\bibinfo{person}{Flask-RESTful Developers}.} \bibinfo{year}{2024}\natexlab{a}.
\newblock \bibinfo{title}{Flask-RESTful Documentation}.
\newblock
\newblock
\urldef\tempurl%
\url{https://flask-restful.readthedocs.io/en/latest/}
\showURL{%
\tempurl}
\newblock
\shownote{Accessed: March 3, 2025}.


\bibitem[Developers(2024b)]%
        {chrome_extensions}
\bibfield{author}{\bibinfo{person}{Google Developers}.} \bibinfo{year}{2024}\natexlab{b}.
\newblock \bibinfo{title}{Chrome Extensions Documentation}.
\newblock
\newblock
\urldef\tempurl%
\url{https://developer.chrome.com/docs/extensions/}
\showURL{%
\tempurl}
\newblock
\shownote{Accessed: March 3, 2025}.


\bibitem[do~Carmo~Nogueira et~al\mbox{.}(2019)]%
        {do2019comparing}
\bibfield{author}{\bibinfo{person}{Tiago do Carmo~Nogueira}, \bibinfo{person}{Deller~James Ferreira}, \bibinfo{person}{S{\'e}rgio~Teixeira de Carvalho}, \bibinfo{person}{Luciana de Oliveira~Berretta}, {and} \bibinfo{person}{Mycke~R Guntijo}.} \bibinfo{year}{2019}\natexlab{}.
\newblock \showarticletitle{Comparing sighted and blind users task performance in responsive and non-responsive web design}.
\newblock \bibinfo{journal}{\emph{Knowledge and Information Systems}}  \bibinfo{volume}{58} (\bibinfo{year}{2019}), \bibinfo{pages}{319--339}.
\newblock


\bibitem[El-Kassas et~al\mbox{.}(2021)]%
        {el2021automatic}
\bibfield{author}{\bibinfo{person}{Wafaa~S El-Kassas}, \bibinfo{person}{Cherif~R Salama}, \bibinfo{person}{Ahmed~A Rafea}, {and} \bibinfo{person}{Hoda~K Mohamed}.} \bibinfo{year}{2021}\natexlab{}.
\newblock \showarticletitle{Automatic text summarization: A comprehensive survey}.
\newblock \bibinfo{journal}{\emph{Expert systems with applications}}  \bibinfo{volume}{165} (\bibinfo{year}{2021}), \bibinfo{pages}{113679}.
\newblock


\bibitem[Fakrudeen et~al\mbox{.}(2017)]%
        {fakrudeen2017finger}
\bibfield{author}{\bibinfo{person}{Mohammed Fakrudeen}, \bibinfo{person}{Sufian Yousef}, {and} \bibinfo{person}{Mahdi~H Miraz}.} \bibinfo{year}{2017}\natexlab{}.
\newblock \showarticletitle{Finger Based Technique (FBT): An Innovative System for Improved Usability for the Blind Users' Dynamic Interaction with Mobile Touch Screen Devices}.
\newblock \bibinfo{journal}{\emph{arXiv preprint arXiv:1708.05073}} (\bibinfo{year}{2017}).
\newblock


\bibitem[Fang(2022)]%
        {Fang2022TheEO}
\bibfield{author}{\bibinfo{person}{Limin Fang}.} \bibinfo{year}{2022}\natexlab{}.
\newblock \showarticletitle{The Effects of Online Review Platforms on Restaurant Revenue, Consumer Learning, and Welfare}.
\newblock \bibinfo{journal}{\emph{Manag. Sci.}}  \bibinfo{volume}{68} (\bibinfo{year}{2022}), \bibinfo{pages}{8116--8143}.
\newblock
\urldef\tempurl%
\url{https://api.semanticscholar.org/CorpusID:246595766}
\showURL{%
\tempurl}


\bibitem[Ferdous et~al\mbox{.}(2022)]%
        {ferdous2022insupport}
\bibfield{author}{\bibinfo{person}{Javedul Ferdous}, \bibinfo{person}{Hae-Na Lee}, \bibinfo{person}{Sampath Jayarathna}, {and} \bibinfo{person}{Vikas Ashok}.} \bibinfo{year}{2022}\natexlab{}.
\newblock \showarticletitle{InSupport: Proxy Interface for Enabling Efficient Non-Visual Interaction with Web Data Records}. In \bibinfo{booktitle}{\emph{27th International Conference on Intelligent User Interfaces}}. \bibinfo{pages}{49--62}.
\newblock


\bibitem[Ferdous et~al\mbox{.}(2021)]%
        {ferdous2021semantic}
\bibfield{author}{\bibinfo{person}{Javedul Ferdous}, \bibinfo{person}{Sami Uddin}, {and} \bibinfo{person}{Vikas Ashok}.} \bibinfo{year}{2021}\natexlab{}.
\newblock \showarticletitle{Semantic table-of-contents for efficient web screen reading}. In \bibinfo{booktitle}{\emph{Proceedings of the 36th Annual ACM Symposium on Applied Computing}}. \bibinfo{pages}{1941--1949}.
\newblock


\bibitem[Filieri(2015)]%
        {filieri2015makes}
\bibfield{author}{\bibinfo{person}{Raffaele Filieri}.} \bibinfo{year}{2015}\natexlab{}.
\newblock \showarticletitle{What makes online reviews helpful? A diagnosticity-adoption framework to explain informational and normative influences in e-WOM}.
\newblock \bibinfo{journal}{\emph{Journal of business research}} \bibinfo{volume}{68}, \bibinfo{number}{6} (\bibinfo{year}{2015}), \bibinfo{pages}{1261--1270}.
\newblock


\bibitem[Foundation(2024)]%
        {python_re}
\bibfield{author}{\bibinfo{person}{Python~Software Foundation}.} \bibinfo{year}{2024}\natexlab{}.
\newblock \bibinfo{title}{re — Regular expression operations}.
\newblock
\newblock
\urldef\tempurl%
\url{https://docs.python.org/3/library/re.html}
\showURL{%
\tempurl}
\newblock
\shownote{Accessed: March 3, 2025}.


\bibitem[{Freedom Scientific}(2022)]%
        {jaws}
\bibfield{author}{\bibinfo{person}{{Freedom Scientific}}.} \bibinfo{year}{2022}\natexlab{}.
\newblock \bibinfo{title}{JAWS \textsuperscript{\textregistered} -- Freedom Scientific}.
\newblock \bibinfo{howpublished}{\url{https://www.freedomscientific.com/products/software/jaws/}}.
\newblock


\bibitem[Fr{\^\i}ncu(2023)]%
        {frincu2023search}
\bibfield{author}{\bibinfo{person}{Ioana Fr{\^\i}ncu}.} \bibinfo{year}{2023}\natexlab{}.
\newblock \showarticletitle{In search of the perfect prompt}.
\newblock  (\bibinfo{year}{2023}).
\newblock


\bibitem[Guerino and Valentim(2020)]%
        {guerino2020usability}
\bibfield{author}{\bibinfo{person}{Guilherme~Corredato Guerino} {and} \bibinfo{person}{Natasha Malveira~Costa Valentim}.} \bibinfo{year}{2020}\natexlab{}.
\newblock \showarticletitle{Usability and user experience evaluation of natural user interfaces: a systematic mapping study}.
\newblock \bibinfo{journal}{\emph{Iet Software}} \bibinfo{volume}{14}, \bibinfo{number}{5} (\bibinfo{year}{2020}), \bibinfo{pages}{451--467}.
\newblock


\bibitem[Hadar-Shoval et~al\mbox{.}(2023)]%
        {hadar2023plasticity}
\bibfield{author}{\bibinfo{person}{Dorit Hadar-Shoval}, \bibinfo{person}{Zohar Elyoseph}, {and} \bibinfo{person}{Maya Lvovsky}.} \bibinfo{year}{2023}\natexlab{}.
\newblock \showarticletitle{The plasticity of ChatGPT’s mentalizing abilities: personalization for personality structures}.
\newblock \bibinfo{journal}{\emph{Frontiers in Psychiatry}}  \bibinfo{volume}{14} (\bibinfo{year}{2023}), \bibinfo{pages}{1234397}.
\newblock


\bibitem[Hart and Staveland(1988)]%
        {hart1998development}
\bibfield{author}{\bibinfo{person}{Sandra~G. Hart} {and} \bibinfo{person}{Lowell~E. Staveland}.} \bibinfo{year}{1988}\natexlab{}.
\newblock \showarticletitle{Development of NASA-TLX (Task Load Index): Results of Empirical and Theoretical Research}.
\newblock In \bibinfo{booktitle}{\emph{Human Mental Workload}}, \bibfield{editor}{\bibinfo{person}{Peter~A. Hancock} {and} \bibinfo{person}{Najmedin Meshkati}} (Eds.). \bibinfo{series}{Advances in Psychology}, Vol.~\bibinfo{volume}{52}. \bibinfo{publisher}{North-Holland}, \bibinfo{pages}{139 -- 183}.
\newblock
\showISSN{0166-4115}
\urldef\tempurl%
\url{https://doi.org/10.1016/S0166-4115(08)62386-9}
\showDOI{\tempurl}


\bibitem[Jin et~al\mbox{.}(2023)]%
        {jin2023binary}
\bibfield{author}{\bibinfo{person}{Xin Jin}, \bibinfo{person}{Jonathan Larson}, \bibinfo{person}{Weiwei Yang}, {and} \bibinfo{person}{Zhiqiang Lin}.} \bibinfo{year}{2023}\natexlab{}.
\newblock \showarticletitle{Binary code summarization: Benchmarking chatgpt/gpt-4 and other large language models}.
\newblock \bibinfo{journal}{\emph{arXiv preprint arXiv:2312.09601}} (\bibinfo{year}{2023}).
\newblock


\bibitem[Karamad(2023)]%
        {Karamad2023THEIO}
\bibfield{author}{\bibinfo{person}{Elnaz Karamad}.} \bibinfo{year}{2023}\natexlab{}.
\newblock \showarticletitle{THE IMPACT OF ONLINE REVIEWS ON CONSUMER DECISION-MAKING: A SURVEY IN THE UKRAINIAN MARKET}.
\newblock \bibinfo{journal}{\emph{Market economy: modern management theory and practice}} (\bibinfo{year}{2023}).
\newblock
\urldef\tempurl%
\url{https://api.semanticscholar.org/CorpusID:268133863}
\showURL{%
\tempurl}


\bibitem[Khalid et~al\mbox{.}(2020)]%
        {khalid2020gbsvm}
\bibfield{author}{\bibinfo{person}{Madiha Khalid}, \bibinfo{person}{Imran Ashraf}, \bibinfo{person}{Arif Mehmood}, \bibinfo{person}{Saleem Ullah}, \bibinfo{person}{Maqsood Ahmad}, {and} \bibinfo{person}{Gyu~Sang Choi}.} \bibinfo{year}{2020}\natexlab{}.
\newblock \showarticletitle{GBSVM: sentiment classification from unstructured reviews using ensemble classifier}.
\newblock \bibinfo{journal}{\emph{Applied Sciences}} \bibinfo{volume}{10}, \bibinfo{number}{8} (\bibinfo{year}{2020}), \bibinfo{pages}{2788}.
\newblock


\bibitem[Khan et~al\mbox{.}(2020)]%
        {khan2020sentiment}
\bibfield{author}{\bibinfo{person}{Aurangzeb Khan}, \bibinfo{person}{Umair Younis}, \bibinfo{person}{Alam~Sher Kundi}, \bibinfo{person}{Muhammad~Zubair Asghar}, \bibinfo{person}{Irfan Ullah}, \bibinfo{person}{Nida Aslam}, {and} \bibinfo{person}{Imran Ahmed}.} \bibinfo{year}{2020}\natexlab{}.
\newblock \showarticletitle{Sentiment classification of user reviews using supervised learning techniques with comparative opinion mining perspective}. In \bibinfo{booktitle}{\emph{Advances in Computer Vision: Proceedings of the 2019 Computer Vision Conference (CVC), Volume 2 1}}. Springer, \bibinfo{pages}{23--29}.
\newblock


\bibitem[Khanna et~al\mbox{.}(2024)]%
        {khanna2024hand}
\bibfield{author}{\bibinfo{person}{Prerna Khanna}, \bibinfo{person}{IV Ramakrishnan}, \bibinfo{person}{Shubham Jain}, \bibinfo{person}{Xiaojun Bi}, {and} \bibinfo{person}{Aruna Balasubramanian}.} \bibinfo{year}{2024}\natexlab{}.
\newblock \showarticletitle{Hand Gesture Recognition for Blind Users by Tracking 3D Gesture Trajectory}. In \bibinfo{booktitle}{\emph{Proceedings of the 2024 CHI Conference on Human Factors in Computing Systems}}. \bibinfo{pages}{1--15}.
\newblock


\bibitem[Kherwa and Bansal(2017)]%
        {kherwa2017latent}
\bibfield{author}{\bibinfo{person}{Pooja Kherwa} {and} \bibinfo{person}{Poonam Bansal}.} \bibinfo{year}{2017}\natexlab{}.
\newblock \showarticletitle{Latent semantic analysis: an approach to understand semantic of text}. In \bibinfo{booktitle}{\emph{2017 International Conference on Current Trends in Computer, Electrical, Electronics and Communication (CTCEEC)}}. IEEE, \bibinfo{pages}{870--874}.
\newblock


\bibitem[Kumar et~al\mbox{.}(2021)]%
        {kumar2021comparing}
\bibfield{author}{\bibinfo{person}{Shashank Kumar}, \bibinfo{person}{Jeevitha Shree~DV}, {and} \bibinfo{person}{Pradipta Biswas}.} \bibinfo{year}{2021}\natexlab{}.
\newblock \showarticletitle{Comparing ten WCAG tools for accessibility evaluation of websites}.
\newblock \bibinfo{journal}{\emph{Technology and Disability}} \bibinfo{volume}{33}, \bibinfo{number}{3} (\bibinfo{year}{2021}), \bibinfo{pages}{163--185}.
\newblock


\bibitem[Lazar et~al\mbox{.}(2007)]%
        {lazar2007frustrates}
\bibfield{author}{\bibinfo{person}{Jonathan Lazar}, \bibinfo{person}{Aaron Allen}, \bibinfo{person}{Jason Kleinman}, {and} \bibinfo{person}{Chris Malarkey}.} \bibinfo{year}{2007}\natexlab{}.
\newblock \showarticletitle{What frustrates screen reader users on the web: A study of 100 blind users}.
\newblock \bibinfo{journal}{\emph{International Journal of human-computer interaction}} \bibinfo{volume}{22}, \bibinfo{number}{3} (\bibinfo{year}{2007}), \bibinfo{pages}{247--269}.
\newblock


\bibitem[Lee and Ashok(2021)]%
        {lee2021towards}
\bibfield{author}{\bibinfo{person}{Hae-Na Lee} {and} \bibinfo{person}{Vikas Ashok}.} \bibinfo{year}{2021}\natexlab{}.
\newblock \showarticletitle{Towards Enhancing Blind Users' Interaction Experience with Online Videos via Motion Gestures}. In \bibinfo{booktitle}{\emph{Proceedings of the 32nd ACM Conference on Hypertext and Social Media}}. \bibinfo{pages}{231--236}.
\newblock


\bibitem[Lee and Ashok(2022)]%
        {lee2022customizable}
\bibfield{author}{\bibinfo{person}{Hae-Na Lee} {and} \bibinfo{person}{Vikas Ashok}.} \bibinfo{year}{2022}\natexlab{}.
\newblock \showarticletitle{Customizable Tabular Access to Web Data Records for Convenient Low-vision Screen Magnifier Interaction}.
\newblock \bibinfo{journal}{\emph{ACM Transactions on Accessible Computing (TACCESS)}} \bibinfo{volume}{15}, \bibinfo{number}{2} (\bibinfo{year}{2022}), \bibinfo{pages}{1--22}.
\newblock


\bibitem[Lee et~al\mbox{.}(2020)]%
        {lee2020rotate}
\bibfield{author}{\bibinfo{person}{Hae-Na Lee}, \bibinfo{person}{Vikas Ashok}, {and} \bibinfo{person}{IV Ramakrishnan}.} \bibinfo{year}{2020}\natexlab{}.
\newblock \showarticletitle{Rotate-and-Press: A Non-visual Alternative to Point-and-Click?}. In \bibinfo{booktitle}{\emph{HCI International 2020--Late Breaking Papers: Universal Access and Inclusive Design: 22nd HCI International Conference, HCII 2020, Copenhagen, Denmark, July 19--24, 2020, Proceedings}}. Springer, \bibinfo{pages}{291--305}.
\newblock


\bibitem[Li et~al\mbox{.}(2023)]%
        {li2023automating}
\bibfield{author}{\bibinfo{person}{Nao Li}, \bibinfo{person}{Xiaoyu Yang}, \bibinfo{person}{IpKin~Anthony Wong}, \bibinfo{person}{Rob Law}, \bibinfo{person}{Jing~Yang Xu}, {and} \bibinfo{person}{Binru Zhang}.} \bibinfo{year}{2023}\natexlab{}.
\newblock \showarticletitle{Automating tourism online reviews: a neural network based aspect-oriented sentiment classification}.
\newblock \bibinfo{journal}{\emph{Journal of Hospitality and Tourism Technology}} \bibinfo{volume}{14}, \bibinfo{number}{1} (\bibinfo{year}{2023}), \bibinfo{pages}{1--20}.
\newblock


\bibitem[Li et~al\mbox{.}(2024)]%
        {li2024guiding}
\bibfield{author}{\bibinfo{person}{Zekun Li}, \bibinfo{person}{Baolin Peng}, \bibinfo{person}{Pengcheng He}, \bibinfo{person}{Michel Galley}, \bibinfo{person}{Jianfeng Gao}, {and} \bibinfo{person}{Xifeng Yan}.} \bibinfo{year}{2024}\natexlab{}.
\newblock \showarticletitle{Guiding large language models via directional stimulus prompting}.
\newblock \bibinfo{journal}{\emph{Advances in Neural Information Processing Systems}}  \bibinfo{volume}{36} (\bibinfo{year}{2024}).
\newblock


\bibitem[Liu(2022)]%
        {liu2022sentiment}
\bibfield{author}{\bibinfo{person}{Bing Liu}.} \bibinfo{year}{2022}\natexlab{}.
\newblock \bibinfo{booktitle}{\emph{Sentiment analysis and opinion mining}}.
\newblock \bibinfo{publisher}{Springer Nature}.
\newblock


\bibitem[Low et~al\mbox{.}(2019)]%
        {low2019twitter}
\bibfield{author}{\bibinfo{person}{Christina Low}, \bibinfo{person}{Emma McCamey}, \bibinfo{person}{Cole Gleason}, \bibinfo{person}{Patrick Carrington}, \bibinfo{person}{Jeffrey~P Bigham}, {and} \bibinfo{person}{Amy Pavel}.} \bibinfo{year}{2019}\natexlab{}.
\newblock \showarticletitle{Twitter a11y: A browser extension to describe images}. In \bibinfo{booktitle}{\emph{Proceedings of the 21st International ACM SIGACCESS Conference on Computers and Accessibility}}. \bibinfo{pages}{551--553}.
\newblock


\bibitem[Luca(2016a)]%
        {luca2016reviews}
\bibfield{author}{\bibinfo{person}{Michael Luca}.} \bibinfo{year}{2016}\natexlab{a}.
\newblock \showarticletitle{Reviews, reputation, and revenue: The case of Yelp. com}.
\newblock \bibinfo{journal}{\emph{Com (March 15, 2016). Harvard Business School NOM Unit Working Paper}} \bibinfo{number}{12-016} (\bibinfo{year}{2016}).
\newblock


\bibitem[Luca(2016b)]%
        {Luca2016ReviewsRA}
\bibfield{author}{\bibinfo{person}{Michael Luca}.} \bibinfo{year}{2016}\natexlab{b}.
\newblock \showarticletitle{Reviews, Reputation, and Revenue: The Case of Yelp.Com}.
\newblock
\urldef\tempurl%
\url{https://api.semanticscholar.org/CorpusID:14511907}
\showURL{%
\tempurl}


\bibitem[Luy et~al\mbox{.}(2021)]%
        {luy2021toolkit}
\bibfield{author}{\bibinfo{person}{Calvin Luy}, \bibinfo{person}{Jeremy Law}, \bibinfo{person}{Lily Ho}, \bibinfo{person}{Richard Matheson}, \bibinfo{person}{Tracey Cai}, \bibinfo{person}{Anuradha Madugalla}, {and} \bibinfo{person}{John Grundy}.} \bibinfo{year}{2021}\natexlab{}.
\newblock \showarticletitle{A toolkit for building more adaptable user interfaces for vision-impaired users}. In \bibinfo{booktitle}{\emph{2021 IEEE symposium on visual languages and human-centric computing (VL/HCC)}}. IEEE, \bibinfo{pages}{1--5}.
\newblock


\bibitem[Melnyk et~al\mbox{.}(2015)]%
        {melnyk2015look}
\bibfield{author}{\bibinfo{person}{Valentyn Melnyk}, \bibinfo{person}{Vikas Ashok}, \bibinfo{person}{Valentyn Melnyk}, \bibinfo{person}{Yury Puzis}, \bibinfo{person}{Yevgen Borodin}, \bibinfo{person}{Andrii Soviak}, {and} \bibinfo{person}{IV Ramakrishnan}.} \bibinfo{year}{2015}\natexlab{}.
\newblock \showarticletitle{Look ma, no aria: generic accessible interfaces for web widgets}. In \bibinfo{booktitle}{\emph{Proceedings of the 12th International Web for All Conference}}. \bibinfo{pages}{1--4}.
\newblock


\bibitem[Nebeling et~al\mbox{.}(2013)]%
        {nebeling2013crowdadapt}
\bibfield{author}{\bibinfo{person}{Michael Nebeling}, \bibinfo{person}{Maximilian Speicher}, {and} \bibinfo{person}{Moira~C Norrie}.} \bibinfo{year}{2013}\natexlab{}.
\newblock \showarticletitle{CrowdAdapt: enabling crowdsourced web page adaptation for individual viewing conditions and preferences}. In \bibinfo{booktitle}{\emph{Proceedings of the 5th ACM SIGCHI symposium on Engineering interactive computing systems}}. \bibinfo{pages}{23--32}.
\newblock


\bibitem[{NV Access}(2022)]%
        {nvaccess}
\bibfield{author}{\bibinfo{person}{{NV Access}}.} \bibinfo{year}{2022}\natexlab{}.
\newblock \bibinfo{title}{NV Access}.
\newblock \bibinfo{howpublished}{\url{https://www.nvaccess.org/}}.
\newblock


\bibitem[Oh et~al\mbox{.}(2021)]%
        {oh2021image}
\bibfield{author}{\bibinfo{person}{Uran Oh}, \bibinfo{person}{Hwayeon Joh}, {and} \bibinfo{person}{YunJung Lee}.} \bibinfo{year}{2021}\natexlab{}.
\newblock \showarticletitle{Image accessibility for screen reader users: A systematic review and a road map}.
\newblock \bibinfo{journal}{\emph{Electronics}} \bibinfo{volume}{10}, \bibinfo{number}{8} (\bibinfo{year}{2021}), \bibinfo{pages}{953}.
\newblock


\bibitem[Oktay(2012)]%
        {oktay2012grounded}
\bibfield{author}{\bibinfo{person}{Julianne~S Oktay}.} \bibinfo{year}{2012}\natexlab{}.
\newblock \bibinfo{booktitle}{\emph{Grounded theory}}.
\newblock \bibinfo{publisher}{Oxford University Press}.
\newblock


\bibitem[OpenAI(2024)]%
        {openai_gpt4}
\bibfield{author}{\bibinfo{person}{OpenAI}.} \bibinfo{year}{2024}\natexlab{}.
\newblock \bibinfo{title}{GPT-4 Model Documentation}.
\newblock
\newblock
\urldef\tempurl%
\url{https://platform.openai.com/docs/models/gpt-4}
\showURL{%
\tempurl}
\newblock
\shownote{Accessed: March 3, 2025}.


\bibitem[Prakash et~al\mbox{.}(2024)]%
        {prakash2024all}
\bibfield{author}{\bibinfo{person}{Yash Prakash}, \bibinfo{person}{Akshay~Kolgar Nayak}, \bibinfo{person}{Mohan Sunkara}, \bibinfo{person}{Sampath Jayarathna}, \bibinfo{person}{Hae-Na Lee}, {and} \bibinfo{person}{Vikas Ashok}.} \bibinfo{year}{2024}\natexlab{}.
\newblock \showarticletitle{All in One Place: Ensuring Usable Access to Online Shopping Items for Blind Users}.
\newblock \bibinfo{journal}{\emph{Proceedings of the ACM on Human-Computer Interaction}} \bibinfo{volume}{8}, \bibinfo{number}{EICS} (\bibinfo{year}{2024}), \bibinfo{pages}{1--25}.
\newblock


\bibitem[Prakash et~al\mbox{.}(2023)]%
        {prakash2023autodesc}
\bibfield{author}{\bibinfo{person}{Yash Prakash}, \bibinfo{person}{Mohan Sunkara}, \bibinfo{person}{Hae-Na Lee}, \bibinfo{person}{Sampath Jayarathna}, {and} \bibinfo{person}{Vikas Ashok}.} \bibinfo{year}{2023}\natexlab{}.
\newblock \showarticletitle{AutoDesc: facilitating convenient perusal of web data items for blind users}. In \bibinfo{booktitle}{\emph{Proceedings of the 28th International Conference on Intelligent User Interfaces}}. \bibinfo{pages}{32--45}.
\newblock


\bibitem[Racherla and Friske(2012)]%
        {Racherla2012PerceivedO}
\bibfield{author}{\bibinfo{person}{Pradeep Racherla} {and} \bibinfo{person}{Wesley Friske}.} \bibinfo{year}{2012}\natexlab{}.
\newblock \showarticletitle{Perceived 'usefulness' of online consumer reviews: An exploratory investigation across three services categories}.
\newblock \bibinfo{journal}{\emph{Electron. Commer. Res. Appl.}}  \bibinfo{volume}{11} (\bibinfo{year}{2012}), \bibinfo{pages}{548--559}.
\newblock
\urldef\tempurl%
\url{https://api.semanticscholar.org/CorpusID:28895996}
\showURL{%
\tempurl}


\bibitem[Salda{\~n}a(2021)]%
        {saldana2021coding}
\bibfield{author}{\bibinfo{person}{Johnny Salda{\~n}a}.} \bibinfo{year}{2021}\natexlab{}.
\newblock \showarticletitle{The coding manual for qualitative researchers}.
\newblock  (\bibinfo{year}{2021}).
\newblock


\bibitem[Sharif et~al\mbox{.}(2021)]%
        {sharif2021understanding}
\bibfield{author}{\bibinfo{person}{Ather Sharif}, \bibinfo{person}{Sanjana~Shivani Chintalapati}, \bibinfo{person}{Jacob~O Wobbrock}, {and} \bibinfo{person}{Katharina Reinecke}.} \bibinfo{year}{2021}\natexlab{}.
\newblock \showarticletitle{Understanding screen-reader users’ experiences with online data visualizations}. In \bibinfo{booktitle}{\emph{Proceedings of the 23rd International ACM SIGACCESS Conference on Computers and Accessibility}}. \bibinfo{pages}{1--16}.
\newblock


\bibitem[Siahaan et~al\mbox{.}(2017)]%
        {siahaan2017new}
\bibfield{author}{\bibinfo{person}{Daniel Siahaan}, \bibinfo{person}{Achmad Maududie}, \bibinfo{person}{Agus Subekti}, \bibinfo{person}{Hotniar Siringoringo}, {et~al\mbox{.}}} \bibinfo{year}{2017}\natexlab{}.
\newblock \showarticletitle{A new approach for modeling researcher community based on scientific article metadata using natural language processing}. In \bibinfo{booktitle}{\emph{2017 Seventh International Conference on Information Science and Technology (ICIST)}}. IEEE, \bibinfo{pages}{386--390}.
\newblock


\bibitem[Sparks and Browning(2011)]%
        {sparks2011impact}
\bibfield{author}{\bibinfo{person}{Beverley~A Sparks} {and} \bibinfo{person}{Victoria Browning}.} \bibinfo{year}{2011}\natexlab{}.
\newblock \showarticletitle{The impact of online reviews on hotel booking intentions and perception of trust}.
\newblock \bibinfo{journal}{\emph{Tourism management}} \bibinfo{volume}{32}, \bibinfo{number}{6} (\bibinfo{year}{2011}), \bibinfo{pages}{1310--1323}.
\newblock


\bibitem[Srinivas et~al\mbox{.}(2024)]%
        {srinivas2024evaluation}
\bibfield{author}{\bibinfo{person}{Manjunath Srinivas}, \bibinfo{person}{S~Vamsi Krishna~Reddy}, \bibinfo{person}{Manoj NM}, {and} \bibinfo{person}{Hatsuho Miyazawa}.} \bibinfo{year}{2024}\natexlab{}.
\newblock \showarticletitle{Evaluation of ChatGPT, Gemini and Llama-2 for E-commerce Product Attribute Extraction}. In \bibinfo{booktitle}{\emph{Proceedings of the 2024 10th International Conference on e-Society, e-Learning and e-Technologies (ICSLT)}}. \bibinfo{pages}{43--48}.
\newblock


\bibitem[Stent et~al\mbox{.}(2005)]%
        {stent2005evaluating}
\bibfield{author}{\bibinfo{person}{Amanda Stent}, \bibinfo{person}{Matthew Marge}, {and} \bibinfo{person}{Mohit Singhai}.} \bibinfo{year}{2005}\natexlab{}.
\newblock \showarticletitle{Evaluating evaluation methods for generation in the presence of variation}. In \bibinfo{booktitle}{\emph{International conference on intelligent text processing and computational linguistics}}. Springer, \bibinfo{pages}{341--351}.
\newblock


\bibitem[Sun et~al\mbox{.}(2023)]%
        {sun-etal-2023-text}
\bibfield{author}{\bibinfo{person}{Xiaofei Sun}, \bibinfo{person}{Xiaoya Li}, \bibinfo{person}{Jiwei Li}, \bibinfo{person}{Fei Wu}, \bibinfo{person}{Shangwei Guo}, \bibinfo{person}{Tianwei Zhang}, {and} \bibinfo{person}{Guoyin Wang}.} \bibinfo{year}{2023}\natexlab{}.
\newblock \showarticletitle{Text Classification via Large Language Models}. In \bibinfo{booktitle}{\emph{Findings of the Association for Computational Linguistics: EMNLP 2023}}, \bibfield{editor}{\bibinfo{person}{Houda Bouamor}, \bibinfo{person}{Juan Pino}, {and} \bibinfo{person}{Kalika Bali}} (Eds.). \bibinfo{publisher}{Association for Computational Linguistics}, \bibinfo{address}{Singapore}, \bibinfo{pages}{8990--9005}.
\newblock
\urldef\tempurl%
\url{https://doi.org/10.18653/v1/2023.findings-emnlp.603}
\showDOI{\tempurl}


\bibitem[Sunkara et~al\mbox{.}(2023)]%
        {sunkara2023enabling}
\bibfield{author}{\bibinfo{person}{Mohan Sunkara}, \bibinfo{person}{Yash Prakash}, \bibinfo{person}{Hae-Na Lee}, \bibinfo{person}{Sampath Jayarathna}, {and} \bibinfo{person}{Vikas Ashok}.} \bibinfo{year}{2023}\natexlab{}.
\newblock \showarticletitle{Enabling customization of discussion forums for blind users}.
\newblock \bibinfo{journal}{\emph{Proceedings of the ACM on Human-Computer Interaction}} \bibinfo{volume}{7}, \bibinfo{number}{EICS} (\bibinfo{year}{2023}), \bibinfo{pages}{1--20}.
\newblock


\bibitem[Tas and Kiyani(2007)]%
        {tas2007survey}
\bibfield{author}{\bibinfo{person}{Oguzhan Tas} {and} \bibinfo{person}{Farzad Kiyani}.} \bibinfo{year}{2007}\natexlab{}.
\newblock \showarticletitle{A survey automatic text summarization}.
\newblock \bibinfo{journal}{\emph{PressAcademia Procedia}} \bibinfo{volume}{5}, \bibinfo{number}{1} (\bibinfo{year}{2007}), \bibinfo{pages}{205--213}.
\newblock


\bibitem[Team et~al\mbox{.}(2023)]%
        {team2023gemini}
\bibfield{author}{\bibinfo{person}{Gemini Team}, \bibinfo{person}{Rohan Anil}, \bibinfo{person}{Sebastian Borgeaud}, \bibinfo{person}{Jean-Baptiste Alayrac}, \bibinfo{person}{Jiahui Yu}, \bibinfo{person}{Radu Soricut}, \bibinfo{person}{Johan Schalkwyk}, \bibinfo{person}{Andrew~M Dai}, \bibinfo{person}{Anja Hauth}, \bibinfo{person}{Katie Millican}, {et~al\mbox{.}}} \bibinfo{year}{2023}\natexlab{}.
\newblock \showarticletitle{Gemini: a family of highly capable multimodal models}.
\newblock \bibinfo{journal}{\emph{arXiv preprint arXiv:2312.11805}} (\bibinfo{year}{2023}).
\newblock


\bibitem[Thelwall(2024)]%
        {thelwall2024chatgpt}
\bibfield{author}{\bibinfo{person}{Mike Thelwall}.} \bibinfo{year}{2024}\natexlab{}.
\newblock \showarticletitle{ChatGPT for complex text evaluation tasks}.
\newblock \bibinfo{journal}{\emph{Journal of the Association for Information Science and Technology}} (\bibinfo{year}{2024}).
\newblock


\bibitem[Touvron et~al\mbox{.}(2023)]%
        {touvron2023llama}
\bibfield{author}{\bibinfo{person}{Hugo Touvron}, \bibinfo{person}{Thibaut Lavril}, \bibinfo{person}{Gautier Izacard}, \bibinfo{person}{Xavier Martinet}, \bibinfo{person}{Marie-Anne Lachaux}, \bibinfo{person}{Timoth{\'e}e Lacroix}, \bibinfo{person}{Baptiste Rozi{\`e}re}, \bibinfo{person}{Naman Goyal}, \bibinfo{person}{Eric Hambro}, \bibinfo{person}{Faisal Azhar}, {et~al\mbox{.}}} \bibinfo{year}{2023}\natexlab{}.
\newblock \showarticletitle{Llama: Open and efficient foundation language models}.
\newblock \bibinfo{journal}{\emph{arXiv preprint arXiv:2302.13971}} (\bibinfo{year}{2023}).
\newblock


\bibitem[Valdivia et~al\mbox{.}(2017)]%
        {valdivia2017sentiment}
\bibfield{author}{\bibinfo{person}{Ana Valdivia}, \bibinfo{person}{M~Victoria Luz{\'o}n}, {and} \bibinfo{person}{Francisco Herrera}.} \bibinfo{year}{2017}\natexlab{}.
\newblock \showarticletitle{Sentiment analysis on tripadvisor: Are there inconsistencies in user reviews?}. In \bibinfo{booktitle}{\emph{Hybrid Artificial Intelligent Systems: 12th International Conference, HAIS 2017, La Rioja, Spain, June 21-23, 2017, Proceedings 12}}. Springer, \bibinfo{pages}{15--25}.
\newblock


\bibitem[Vandic et~al\mbox{.}(2024)]%
        {vandic2024framework}
\bibfield{author}{\bibinfo{person}{Damir Vandic}, \bibinfo{person}{Lennart~J Nederstigt}, \bibinfo{person}{Flavius Frasincar}, \bibinfo{person}{Uzay Kaymak}, {and} \bibinfo{person}{Enzo Ido}.} \bibinfo{year}{2024}\natexlab{}.
\newblock \showarticletitle{A framework for approximate product search using faceted navigation and user preference ranking}.
\newblock \bibinfo{journal}{\emph{Data \& Knowledge Engineering}}  \bibinfo{volume}{149} (\bibinfo{year}{2024}), \bibinfo{pages}{102241}.
\newblock


\bibitem[Wallace et~al\mbox{.}(2021)]%
        {wallace2021generating}
\bibfield{author}{\bibinfo{person}{Byron~C Wallace}, \bibinfo{person}{Sayantan Saha}, \bibinfo{person}{Frank Soboczenski}, {and} \bibinfo{person}{Iain~J Marshall}.} \bibinfo{year}{2021}\natexlab{}.
\newblock \showarticletitle{Generating (factual?) narrative summaries of rcts: Experiments with neural multi-document summarization}.
\newblock \bibinfo{journal}{\emph{AMIA Summits on Translational Science Proceedings}}  \bibinfo{volume}{2021} (\bibinfo{year}{2021}), \bibinfo{pages}{605}.
\newblock


\bibitem[{WebAIM}(2019)]%
        {webaim2019screen}
\bibfield{author}{\bibinfo{person}{{WebAIM}}.} \bibinfo{year}{2019}\natexlab{}.
\newblock \bibinfo{title}{WebAIM: Screen Reader User Survey \#8 Results}.
\newblock
\newblock
\urldef\tempurl%
\url{https://webaim.org/projects/screenreadersurvey8/}
\showURL{%
\tempurl}


\bibitem[WebAIM(2015)]%
        {webaim2015screen}
\bibfield{author}{\bibinfo{person}{Jared~Smith WebAIM}.} \bibinfo{year}{2015}\natexlab{}.
\newblock \showarticletitle{Screen reader user survey\# 5 results}.
\newblock \bibinfo{journal}{\emph{http://webaim. org/projects/screenreadersurvey6/.}} (\bibinfo{year}{2015}).
\newblock


\bibitem[{World Wide Web Consortium (W3C)}(2021)]%
        {w3c-aria}
\bibfield{author}{\bibinfo{person}{{World Wide Web Consortium (W3C)}}.} \bibinfo{year}{2021}\natexlab{}.
\newblock \bibinfo{title}{{Accessible Rich Internet Applications (WAI-ARIA) 1.2}}.
\newblock \bibinfo{howpublished}{\url{https://www.w3.org/TR/wai-aria-1.2/}}.
\newblock
\newblock
\shownote{Accessed: 2025-02-14}.


\bibitem[Wu et~al\mbox{.}(2017)]%
        {wu2017automatic}
\bibfield{author}{\bibinfo{person}{Shaomei Wu}, \bibinfo{person}{Jeffrey Wieland}, \bibinfo{person}{Omid Farivar}, {and} \bibinfo{person}{Julie Schiller}.} \bibinfo{year}{2017}\natexlab{}.
\newblock \showarticletitle{Automatic alt-text: Computer-generated image descriptions for blind users on a social network service}. In \bibinfo{booktitle}{\emph{proceedings of the 2017 ACM conference on computer supported cooperative work and social computing}}. \bibinfo{pages}{1180--1192}.
\newblock


\bibitem[Xu et~al\mbox{.}(2012)]%
        {xu2012personalized}
\bibfield{author}{\bibinfo{person}{Jingnan Xu}, \bibinfo{person}{Xiaolin Zheng}, {and} \bibinfo{person}{Weifeng Ding}.} \bibinfo{year}{2012}\natexlab{}.
\newblock \showarticletitle{Personalized recommendation based on reviews and ratings alleviating the sparsity problem of collaborative filtering}. In \bibinfo{booktitle}{\emph{2012 IEEE Ninth International Conference on e-Business Engineering}}. IEEE, \bibinfo{pages}{9--16}.
\newblock


\bibitem[Yang et~al\mbox{.}(2023)]%
        {yang2023exploring}
\bibfield{author}{\bibinfo{person}{Xianjun Yang}, \bibinfo{person}{Yan Li}, \bibinfo{person}{Xinlu Zhang}, \bibinfo{person}{Haifeng Chen}, {and} \bibinfo{person}{Wei Cheng}.} \bibinfo{year}{2023}\natexlab{}.
\newblock \showarticletitle{Exploring the limits of chatgpt for query or aspect-based text summarization}.
\newblock \bibinfo{journal}{\emph{arXiv preprint arXiv:2302.08081}} (\bibinfo{year}{2023}).
\newblock


\bibitem[Yayli and Bayram(2012)]%
        {Yayli2012eWOMTE}
\bibfield{author}{\bibinfo{person}{Ali Yayli} {and} \bibinfo{person}{Murat Bayram}.} \bibinfo{year}{2012}\natexlab{}.
\newblock \showarticletitle{e-WOM: the effects of online consumer reviews on purchasing decisions}.
\newblock \bibinfo{journal}{\emph{International Journal of Internet Marketing and Advertising}}  \bibinfo{volume}{7} (\bibinfo{year}{2012}), \bibinfo{pages}{51}.
\newblock
\urldef\tempurl%
\url{https://api.semanticscholar.org/CorpusID:168100918}
\showURL{%
\tempurl}


\bibitem[Zong et~al\mbox{.}(2022)]%
        {zong2022rich}
\bibfield{author}{\bibinfo{person}{Jonathan Zong}, \bibinfo{person}{Crystal Lee}, \bibinfo{person}{Alan Lundgard}, \bibinfo{person}{JiWoong Jang}, \bibinfo{person}{Daniel Hajas}, {and} \bibinfo{person}{Arvind Satyanarayan}.} \bibinfo{year}{2022}\natexlab{}.
\newblock \showarticletitle{Rich screen reader experiences for accessible data visualization}. In \bibinfo{booktitle}{\emph{Computer Graphics Forum}}, Vol.~\bibinfo{volume}{41}. Wiley Online Library, \bibinfo{pages}{15--27}.
\newblock


\bibitem[Zuheros et~al\mbox{.}(2021)]%
        {zuheros2021sentiment}
\bibfield{author}{\bibinfo{person}{Cristina Zuheros}, \bibinfo{person}{Eugenio Mart{\'\i}nez-C{\'a}mara}, \bibinfo{person}{Enrique Herrera-Viedma}, {and} \bibinfo{person}{Francisco Herrera}.} \bibinfo{year}{2021}\natexlab{}.
\newblock \showarticletitle{Sentiment analysis based multi-person multi-criteria decision making methodology using natural language processing and deep learning for smarter decision aid. Case study of restaurant choice using TripAdvisor reviews}.
\newblock \bibinfo{journal}{\emph{Information Fusion}}  \bibinfo{volume}{68} (\bibinfo{year}{2021}), \bibinfo{pages}{22--36}.
\newblock


\end{thebibliography}
